\documentclass[12pt]{jpconf}

\usepackage{iopams}
\expandafter\let\csname equation*\endcsname\relax
\expandafter\let\csname endequation*\endcsname\relax
\usepackage{graphicx}
\usepackage[latin1]{inputenc}
\usepackage{epsfig}
\usepackage{amsmath}
\usepackage{amsfonts}
\usepackage{float}
\usepackage{amssymb}
\usepackage{color}
\newcommand{\bmat}{\left(\begin{array}}
    \newcommand{\emat}{\end{array}\right)}
\newcommand{\be}{\begin{equation}}
\newcommand{\ee}{\end{equation}}
\newcommand{\bea}{\begin{eqnarray}}
\newcommand{\eea}{\end{eqnarray}}
\newcommand{\ba}{\begin{array}}
    \newcommand{\ea}{\end{array}}
\usepackage[colorlinks,citecolor=blue,urlcolor=blue,linkcolor=blue]{hyperref}

\def\lsim{\raise0.3ex\hbox{$\;<$\kern-0.75em\raise-1.1ex\hbox{$\sim\;$}}}
\def\gsim{\raise0.3ex\hbox{$\;>$\kern-0.75em\raise-1.1ex\hbox{$\sim\;$}}}

\begin{document}

\title{Beyond SM Physics and searches for SUSY at the LHC}

\author{Dris Boubaa$^{1,2}$ and Gaber Faisel$^3$ and Shaaban Khalil$^4$}

\address{$^1$Laboratoire de Physique des Particules et Physique
Statistique, Ecole Normale Sup\'erieure-Kouba,
B.P. 92, 16050, Vieux-Kouba, Algiers, Algeria\\

$^2$Department of Physics, Faculty of Exact Sciences and
Computing,
Hassiba Benbouali University of Chlef, B.P 78C , Ouled Fares Chlef 02180, Algeria\\
$^3$Department of Physics, Faculty of Arts and Sciences, S\"{u}leyman Demirel University, \\
$\hphantom{^1}$Isparta 32260, Turkey\\
$^4$Center  for  Fundamental  Physics, Zewail  City  of  Science
and  Technology, Giza 12588,  Egypt}

\ead{d.boubaa@univ-chlef.dz, gaberfaisel@sdu.edu.tr,
skhalil@zewailcity.edu.eg}

\vspace{10pt}

\begin{abstract}
This is the written version of a talk given by S.K. at the
$10^{th}$ International Conference on High Energy and
Astroparticle, Constantine, Algeria. We briefly review the
Standard Model (SM) and the major evidences and main direction of
physics beyond the SM (BSM). We introduce supersymmetry, as one of
the well-motivated BSM. Basic introduction to Minimal
Supersymmetric Standard Model (MSSM) is given. We analyze the
thermal relic abundance of lightest neutralino, which is the
Lightest Supersymmetric Particle (LSP) in the MSSM. We show that
the combined Large Hadron Collider (LHC) and relic abundance
constraints rule out most of the MSSM parameter space except a
very narrow region.  We also review non-minimal SUSY model, based
on the gauge group $SU(3)_C \times SU(2)_L \times U(1)_Y \times
U(1)_{B-L}$ (BLSSM), where an Inverse Seesaw mechanism of light
neutrino mass generation is naturally implemented. The
phenomenological implications of this type of model at the Large
Hadron Collider (LHC) are analyzed.

\end{abstract}

\section{Introduction}
\vskip 0.2cm
The standard model (SM) of particle physics is proved to be in an
excellent agreement with most of the confirmed experimental
results. For instances, the success of the SM includes, the
discovery of the vector bosons $W$ and $Z$, with masses and decay
properties coincide accurately with the SM expectations. While we
have now an impressive list of experimental confirmations of the
success of the SM, compelling arguments indicate that the SM
cannot be the complete theory of Nature. Among the theoretical
problems that face the SM and strongly suggest New Physics (NP)
beyond the SM are the following. The SM does not include gravity,
therefore it cannot be valid at energy scales above $M_{Pl} \sim
10^{19}$ GeV. Also, the SM does not allow for neutrino masses,
therefore it cannot be even valid at energy scales above $M_{\rm
seesaw} \gsim 1$ TeV. Moreover, the SM fails to address other
issues such as the naturalness problems of the Higgs sector in the
SM, the strength of the charge conjugation-parity (CP) violation
in the SM, which is not sufficient to account for the cosmological
baryon asymmetry of the universe,  and the absence of viable Dark
Matter (DM) candidate in the SM. Therefore, it is common a tempt
to conclude that the SM is only an effective low energy limit of a
more fundamental underlying theory. However, it is mandatory for
any new fundamental theory to exactly reproduce the SM at the
Fermi scale.

The supersymmetric theories are considered as the most promising
candidate for the unified theory beyond the SM. Despite the
absence of experimental verification, relevant theoretical
arguments can be given in favor of supersymmetry (SUSY). First of
all, supersymmetry ensures the stability of hierarchy between the
weak and the Planck scales. If we believe that SM should be
embedded within a more fundamental theory including gravity with a
characteristic scale of order the Planck mass $M_P$, then we are
faced with the hierarchy problem. There is no symmetry protecting
the masses of the scalar particles against quadratic divergences
in the perturbation theory. This problem of stabilizing the scalar
masses is solved in SUSY models since now the scalar mass and the
mass of its superpartner are related. SUSY is also a necessary
ingredient in string theories to avoid tachyons in the spectrum.
Further, the evolution of the three gauge coupling constant of the
SM into a single unification scale, only after taking into account
SUSY as we will show below, is considered as a hint that SUSY
might be true.

 In supersymmetric theories, each particle must have a
superpartner (a boson for a fermion and vice versa). Hence, many
new particles must be included into the theory. It is worthwhile
to mention that some physicist find these plethora of new
particles is a defect of the supersymmetric theory. However, we
would like to remind that there are several reasons to accept the
idea of existing new particles (even the non-baryonic particles).
For instance, in astronomy there is over whelming evidence that
most of the mass in the universe is some non-luminous (and
non-baryonic) DM of as yet unknown composition. The lightest SUSY
particle (which is absolutely stable) is a natural candidate for
solving the DM problem. Moreover, we don't believe that SM
spectrum ( the quarks, leptons, Higgs and gauge bosons) is the
complete list of the elementary particles in nature. We believe
that increasing the energy in the accelerators will give us the
chance to discover new particles. Few decades ago we were only
familiar with less than one-third of the SM particles. Hence, the
existence of new particles with masses larger than the Fermi scale
could be a prediction, may be verified experimentally over the
next years.

It is the aim of this paper to review the Supersymmetric
extensions of the SM. The plan of the paper is as follows. In the
next section we give a brief introduction to the SM. In Sect. 3 we
discus the drawbacks of the SM. Potential directions for physics
beyond the SM are described in Sect. 4. In Sect. 5 Supersymmetry
is introduced as one of the best candidates for physics beyond SM.
Supersymmetric DM is analyzed in Sect. 6. In Sect. 7 we review
non-minimal supersymmetric model that accounts for neutrino
masses. In Sect. 8 we show that in this class of models, the
lightest right-handed sneutrino is an interesting candidate for
scalar DM. We conclude in Sect. 9.

\section{The Standard Model}
\vskip 0.2cm

Most of the available experimental data, up to date, can be
explained well in the framework of the standard model
(SM)\cite{Weinberg:tq,Glashow:tr,Salam}. Not only this, but the SM
successfully  predicted the existence of the third generation of
fermions, scalar boson (Higgs) and massive gauge bosons namely,
$W^\pm, Z^0$ long time even before being discovered at colliders.
The SM is based on a gauge theory describing three fundamental
forces in our universe.  These are, the electromagnetic force, the
weak nuclear force, and the strong nuclear force.  Unfortunately,
gravity can not be  included in this theory. The theory is based
on the gauge symmetry group $SU(3)_C \times SU(2)_L \times
U(1)_Y$, where $C$ represents the color, $L$ denotes left-handed
chirality and $Y$ stands for hypercharge. Gauge bosons are
associated with each gauge symmetry group. These bosons are listed
as:

\begin{equation}
SU(3)_{\mathrm C}\rightarrow 8~ G_{\mu}^{\alpha}(\alpha=1,..,8),
\end{equation}
\begin{equation}
SU(2)_{\mathrm L}\rightarrow 3~ W_{\mu}^{a}(a=1,2,3),
\end{equation}
\begin{equation}
U(1)_{\mathrm Y}\rightarrow B_{\mu}.
\end{equation}

The fermionic sector of the SM includes quarks and leptons ordered
in three families of left-handed doublets and right-handed
singlets. In flavor space they can be written as:
\begin{equation*}\begin{array}{ccccc}
L_1=\left(%
\begin{array}{c}
\nu_e \\
e^- \\
\end{array}%
\right)_L,\, & e_{R_1}=e_R^-,\, & Q_1=\left(%
\begin{array}{c}
u \\
d \\
\end{array}%
\right)_L,\, & u_{R_1}=u_R,\,& d_{R_1}=d_R, \\ \\
L_2=\left(%
\begin{array}{c}
\nu_\mu \\
\mu^- \\
\end{array}%
\right)_L, & e_{R_2}=\mu_R^-, & Q_2=\left(%
\begin{array}{c}
c \\
s \\
\end{array}%
\right)_L, & u_{R_2}=c_R, & d_{R_2}=s_R, \\ \\
L_3=\left(%
\begin{array}{c}
\nu_\tau \\
\tau^- \\
\end{array}%
\right)_L, & e_{R_3}=\tau_R^-, & Q_3=\left(%
\begin{array}{c}
t \\
b \\
\end{array}%
\right)_L, & u_{R_3}=t_R, & d_{R_3}=b_R. \\
\end{array}
\end{equation*}

 The SM Lagrangian has the form
\begin{eqnarray} %
    \label{smlagrangian}%
    {\cal L}_{\rm SM}\!\!\!&\!\!\!=\!\!\!&\!\!\! -\frac{1}{4} G_{\mu \nu}^a G^{\mu \nu}_a
    -\frac{1}{4} F_{\mu \nu}^a F^{\mu \nu}_a -\frac{1}{4}
    B_{\mu \nu}B^{\mu \nu} +  \bar{L_i}\, i D_\mu \gamma^\mu \, L_i  \nonumber\\
    \!\!\!\!&\!\!\!+\!\!\!&\!\!\! \bar{e}_{Ri} \, i
    D_\mu \gamma^\mu \, e_{R_i} \ + \bar{Q_i}\, i D_\mu \gamma^\mu \,
    Q_i +  \bar{u}_{Ri} \, i D_\mu \gamma^\mu \, u_{R_i}
    \ + \bar{d}_{Ri} \, i D_\mu \gamma^\mu \, d_{R_i},
\end{eqnarray}
where the covariant derivatives $D_\mu$ are defined as follows:
\begin{eqnarray} %
    D_{\mu} Q_i &=&  \left(
    \partial_\mu -ig_s T_a G^a_\mu -ig_2 \frac{\tau_a}{2} W^a_\mu -i g_1
    \frac{Y_{Q_i}}{2} B_\mu \right) Q_i ,\\
    D_{\mu} L_i &=& \left(
    \partial_\mu -ig_2 \frac{\tau_a}{2} W^a_\mu -i g_1
    \frac{Y_{L_i}}{2} B_\mu \right) \ L_i , ~~~ D_{\mu} f_R = \left( \partial_\mu -i g_1
    \frac{Y_{f_R}}{2} B_\mu \right) \ f_R,
\end{eqnarray}
with $T_a$, $\tau_a$, $Y$, and $g$'s  represent the generators and
the couplings of the corresponding gauge symmetry groups. The
hypercharge quantum number $Y$ is related to the electric charge,
$Q$, and the third component of the isospin, $I_{3}$, via the
relation $Q=I_{3}+\frac{Y}{2}$.

The scalar sector of the SM consists of only one complex $SU(2)_L$
doublet $\Phi$, known as the Higgs field, with hypercharge $Y=1$
\begin{align}
\Phi &= \begin{pmatrix}
\phi^{+}\\
\phi^{0}
\end{pmatrix},
\end{align}
and can be described by the Lagrangian
\begin{equation}
{\cal L}_{scalar}=(D_\mu \Phi)^\dagger (D^\mu \Phi) -V(\Phi),
\end{equation}

here $V(\Phi)$ is the Higgs potential and can be defined as

\begin{equation}\label{scalarpot}
V(\Phi)=-\mu^{2}\Phi^{\dagger}\Phi+\lambda(\Phi^{\dagger}\Phi)^{2}.
\end{equation}
The parameters $\lambda$ and $\mu^{2}$ must be positive to ensure
that the potential is bounded from below. In this case the
potential has a minimum
\begin{align}
\langle\Phi\rangle_0 &=\langle0|\Phi|0\rangle =\begin{pmatrix}
0\\
\frac{v}{\sqrt{2}}
\end{pmatrix},
\end{align}
where $v/\sqrt{2}$ is the vacuum expectation value (VEV), with
$v=\sqrt{\mu^2/\lambda}$. Only the neutral component of the Higgs
doublet can get a VEV so that  $SU(2)_L$ symmetry is broken while
other symmetries remain unbroken i.e. $ SU(2)_{L}\otimes
U(1)_{Y}\to U(1)_{\rm Q}$. As a consequence, the eight gluons  and
the photon  remain massless while the gauge bosons, $W^\pm$ and
$Z^0$, associated with the $ SU(2)_{L}$, symmetry acquire masses
\cite{Englert:1964et,Higgs:1964pj,Guralnik:1964eu} given as

\begin{equation}
M_{W}^{2}=\frac{1}{4}g^{2}_2v^{2},~~~M_{Z}^{2}=\frac{1}{4}(g^{2}_1+{g}^{2}_2)v^{2},
\end{equation}

\begin{figure}[h]
\begin{center}
\vskip -0.5cm
\includegraphics[height=2cm,width=3 cm]{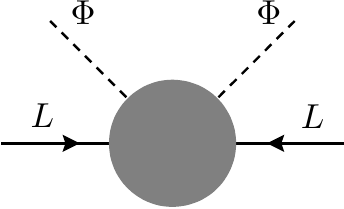}~~~~~~~~~~~   \includegraphics[height=5cm,width=6 cm]{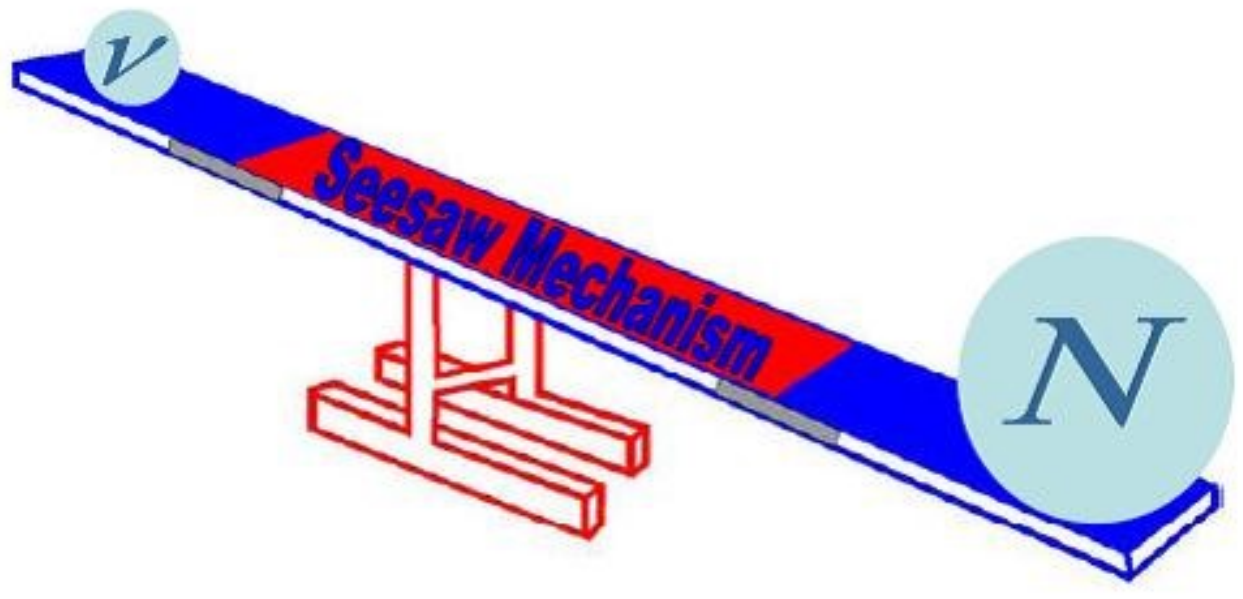}
\caption{ Left: dimension five operator responsible for generating
neutrino mass \cite{Escrihuela:2015wra}. Right: neutrino mass
generations through seesaw mechanism.\label{seesaw}}
\end{center}
\end{figure}

The fermion masses can be also generated via the Yukawa
interaction of the left and right-handed fermion fields to the
Higgs field. The corresponding Lagrangian is given as
\begin{equation}
\mathcal{L}_{ _{Y}}=Y_e^{ij}~ \bar{L}_{i}~\Phi~ e_{Rj}+Y_u^{ij}~
\bar{Q}_{i}~\tilde{\Phi} ~u_{Rj}+Y^{ij}_d~ \bar{Q}_{i}~\Phi
~d_{Rj}+h.c, \label{Lyu}\end{equation} where
$\tilde{\Phi}=i\tau_2\Phi^*$, $Y_{\tilde{\Phi}}=-1$, and $Y^{ij}$
are the Yukawa couplings. After electroweak symmetry breaking, the
Higgs field $\Phi$ can be expressed as,
\begin{align}
\Phi=\begin{pmatrix}
0\\
(v+ h)/\sqrt{2}
\end{pmatrix},
\label{hig}\end{align}

after substitution in Eq.(\ref{Lyu}), one gets the following
fermion mass matrices
\begin{equation}
M_r^{ij}=\frac{v}{\sqrt{2}}Y_r^{ij}\ \ ,\ \ r=e,u,d.
\end{equation}
To obtain the fermion masses, we need to
diagonalize the mass matrices via unitary fields transformations, this simplifies to:
\begin{equation}
m_e=\frac{v}{\sqrt{2}}Y_e,~~~~m_u=\frac{v}{\sqrt{2}}Y_u,~~~~m_d=\frac{v}{\sqrt{2}}Y_d.
\label{yuk}\end{equation} Clearly, only  quarks and charged
leptons electrons acquire masses while neutrinos remain massless
since they do not have right-handed components. As can be seen
From Eqs.(\ref{Lyu} and \ref{hig}), the neutral higss boson, $h$,
interactions with quarks and charged leptons can lead to
phenomenological predictions that can be tested in colliders.
These are also predictions of the SM and thus can be used also as
probes for physics beyond the SM.

\section{Evidence for Physics Beyond the SM}

After discovering the only scalar boson of the SM in 2012, at the
CERN Large Hadron Collider (LHC) by CMS \cite{Chatrchyan:2012xdj}
and ATLAS \cite{Aad:2013wqa}, the SM has been confirmed as being
extremely successful in describing the most aspects of the nature
with remarkable precision at the 100 GeV scale.  However, even
with this success, there are number of theoretical and
phenomenological outstanding issues in the particle physics that
can not be explained and the SM fails to address them adequately.
We discuss some of these issues in the following.

\begin{figure}[h]
\begin{center}
\vskip -0.5cm
\includegraphics[scale=0.5]{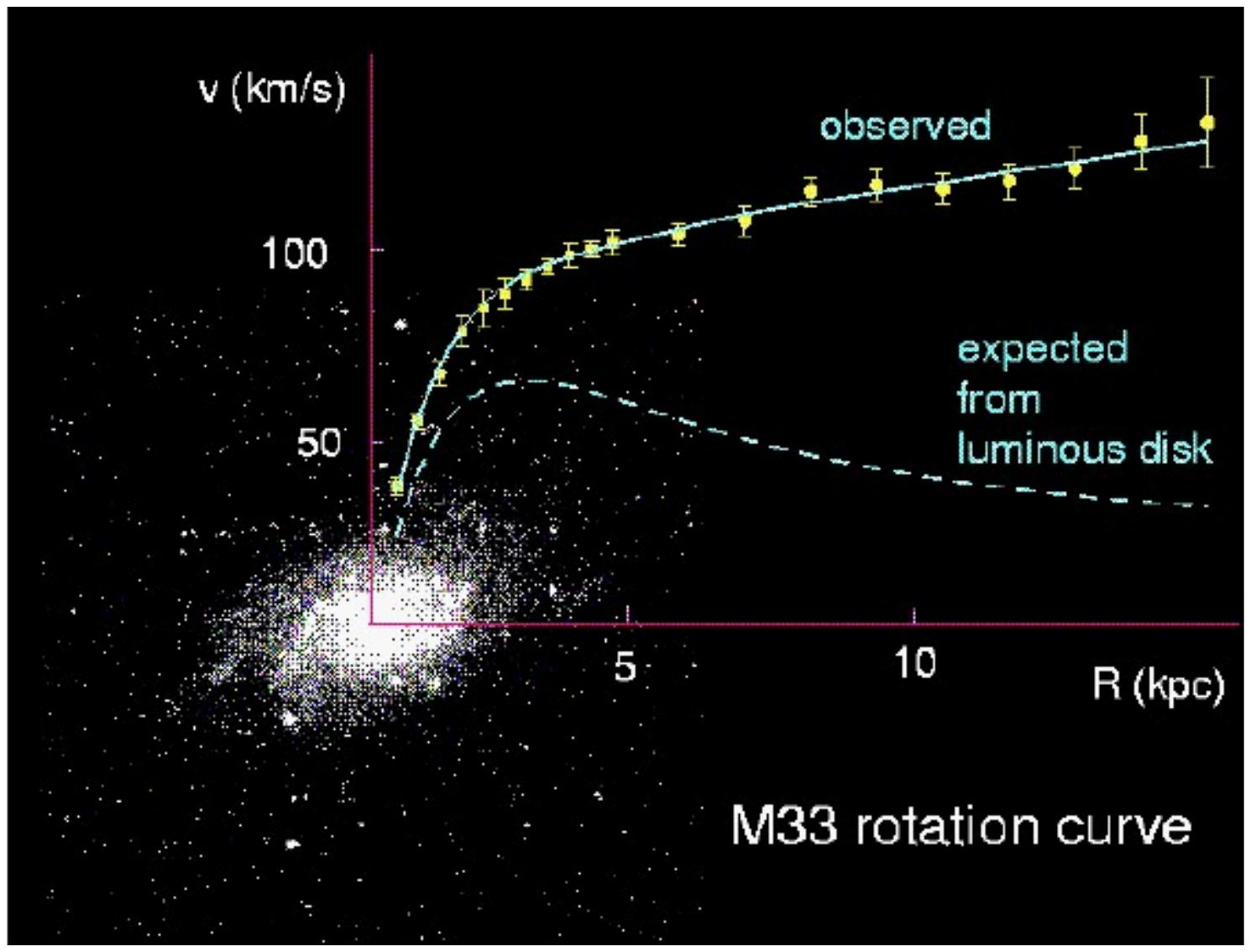}%
\vskip -1.5cm \caption{ The observed rotation curve of the dwarf
spiral galaxy M33 superimposed on its optical image
\cite{Corbelli:1999af,Roy:2000hz}.\label{rot}}
\end{center}
\end{figure}

 In the SM, quarks and electrons acquire masses through Yukawa
couplings as can be seen in Eq.(\ref{yuk}). Neutrinos remain
massless because there are no RH $\nu$ in the SM. However, it has
proven experimentally that $m_{\nu} \neq 0$
\cite{Tanabashi:2018oca}.  Neutrino masses can be generated if
lepton number is violated by dimension 5 operator
\cite{Weinberg:1979sa}. In the literature several mechanisms
mechanism were proposed to generate neutrino masses. For
instances, seesaw mechanism, shown in Fig.\ref{seesaw}, with three
different types\cite{seesaw1,seesaw12,seesaw2,seesaw3}.

In 1933, Zwicky noticed that the mass of luminous matter in the
Coma cluster is much smaller than its total mass
\cite{Zwicky:1933gu}. On the other hand, the observation of 1000
spiral galaxies showed that away from the center of galaxies the
rotation velocities do not drop off with distance as shown in
Fig.\ref{rot} \cite{Corbelli:1999af,Roy:2000hz}. This observation
is  in contradiction with what we expect as the  velocity of
rotating objects is given by $v(r)= \sqrt{\frac{G\ M(r)}{r}}$.
Dark matter (DM) was proposed as a possible explanation for this
observation where the disk galaxies are assumed to be immersed in
extended DM halos. However, in the SM, there is no candidate to
play the role of the cold dark matter of the universe.

\begin{figure}[t]
\begin{center}
\includegraphics[height=6.cm,width=9 cm]{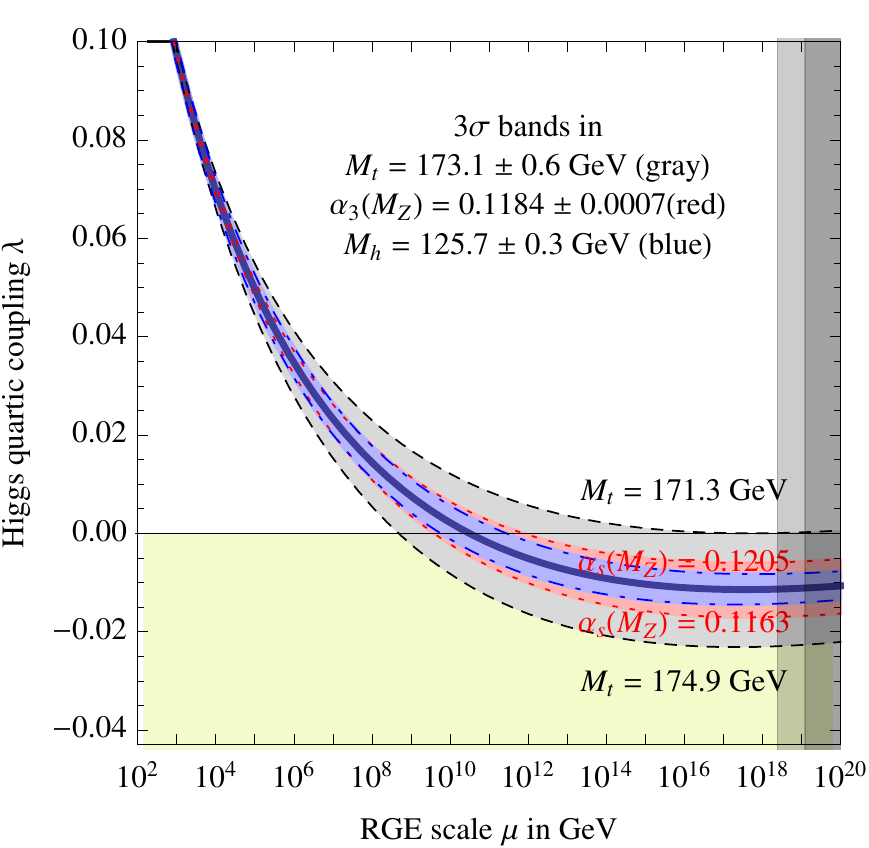}
\caption{ Higgs qadratic coupling running in the SM
\cite{Degrassi:2012ry}.\label{lambda}}
\end{center}
\end{figure}

Other issues also include Higgs vacuum stability where qadratic
coupling evolves to zero or negative values as one can see from
Fig.(\ref{lambda}) \cite{Degrassi:2012ry}. This turns to be a
problem as in the SM $M_H = \sqrt{\lambda} v$. Higgs Mass
Hierarchy is also one of the problems in the SM due to the a
absence of a symmetry to protect Higgs mass that receive
contributions from the loop dominated by top quark contributions.
The contributions are proportional to the square of $\Lambda$, the
cutoff scale of the theory, that can be set to Planck scale
($10^{19}$ GeV).

It is believed that the early universe began as a huge burst of
energy known as the big bang. In the short moment after that bang,
matter and anti-matter existed in that early phase of universe in
equal amounts (CP symmetry). However, with time evolving and
cooling of the universe,  our present universe is dominated by
only matter not their anti-matter. This matter asymmetry of the
universe is known as Baryon Asymmetry (Matter- Antimatter
Asymmetry), cannot be explained in the context of the SM since the
amount of CP violation in the SM is not enough for such
explanation. Neither the standard model of particle physics, nor
the theory of general relativity provides an obvious explanation.
 In 1967, A. Sakharov showed that the generation of the net
baryon number in the universe requires:  Baryon number violation,
thermal non-equilibrium  and C and CP violation
\cite{Sakharov:1967dj}. All of these ingredients were present  in
the early Universe! Do we understand the cause of CP violation in
particle interactions? Can we calculate the BAU from first
principles?
$$ \frac{(n_B -n_{\bar{B}})}{n_\gamma} =6.1 \times 10^{-10}.$$
With all issues discussed above, there are also a number of
questions we hope will be answered in beyond SM physics:

i) Electroweak symmetry breaking, which is not explained within
the SM.

ii) Why is the symmetry group is $SU(3) \times  SU(2) \times
U(1)$?

 iii) Can forces be unified?

 iv) Why are there three families of quarks and leptons?

v) Why do the quarks and leptons have the masses they do?

 vi) Can we have a quantum theory of gravity?

vii)  Why is the cosmological constant much smaller than simple
estimates would suggest?

Based on the above discussion we can conclude that, Standard Model
is defined by 4-dimension QFT (Invariant under Poincare group).
The symmetry governs the SM is local $SU(3)_C \times SU(2)_L
\times U(1)_Y$ and the particle content (Point particles): 3
fermion (quark and Lepton) Generations with no Right-handed
neutrinos resulting in massless neutrinos. In the SM, symmetry
breaking can be achieved via one Higgs doublet.  Moreover, in the
SM, no candidate for Dark Matter and formalism of the SM does not
allow to include gravity.

\section{Directions Beyond the Standard Model}
\vskip 0.2cm
 The failure of the SM to address the problems discussed above
motivates going beyond SM. Directions Beyond the Standard Model
can be achieved upon extending some sectors or symmetries in the
SM in several ways. These directions can be summaries as follows:
\begin{itemize}
\item Extension of gauge symmetry

\item Extension of Higgs Sector

\item Extension of matter content

\item Extension with flavor symmetry

\item Extension of Space-time dimenstions (Extra-dimensions)

\item Extension of Lorentz Symmetry (Supersymmetry)

\item Incorporate Gravity (Supergravity)

\item Adopt the concept of one dimension object, instead of zero-point particle. (Superstring).
\end{itemize}

One of the most popular extension of the SM is Supersymmetry
(SUSY) which is based on a symmetry linking fermions and bosons
and thus enlarge the usual space-time to include fermionic
components. Beyond the SM physics include also string theory and
extra dimensions where the dimensionality of the space-time
increase to include extra dimensions which result in consequence
that can not be seen else where.

\section{Supersymmetry}
\vskip 0.2cm

\begin{figure}[t]
\begin{center}
\vskip -0.5cm
\includegraphics[scale=0.7]{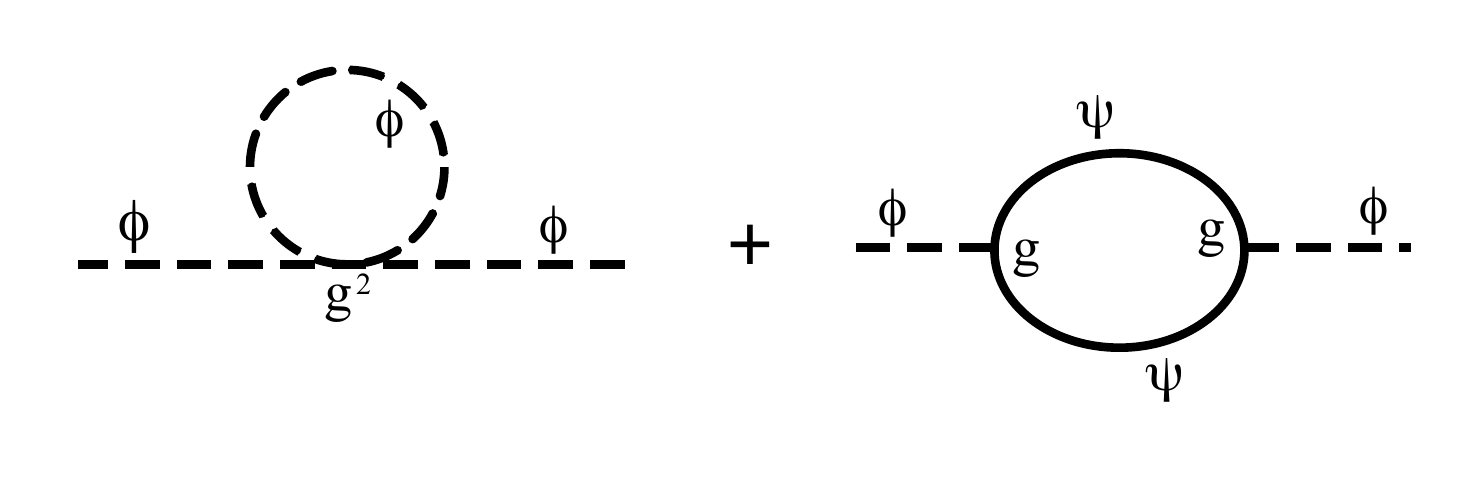}%
\vskip -0.5cm \caption{One loop diagrams which yield a corrections
to the scalar mass \cite{Quevedo:2010ui}.\label{canc}}
\end{center}
\end{figure}

Historically, Supersymmetry was introduced in 1973 as a part of an
extension of the special relativity. Supersymmetric theories are
promising candidates for unified theory beyond the SM. Moreover,
in SUSY, the mechanism of the electroweak symmetry breaking is
natural. Supersymmetry is an extension of the space time symmetry
that relates  bosons and  fermions. As a consequence,
contributions from the new scalar bosons can ensure the stability
of hierarchy between the weak and the Planck scales through
avoiding the fine tuning in the renormalization of the Higgs boson
mass at the level of ${\cal O}(10^{34})$.

Hierarchy problem is one of the naturalness problems of the Higgs
sector in the SM. In the SM, the Higgs mass receives contribution
from one-loop radiative corrections. This contribution is
proportional to the square of the momentum running in the loop and
can be set to the cut-off scale or larger than that. With the
cut-off scale of the order of the GUT scale,
$M_G\approx10^{16}~\mathrm {GeV}$, Higgs mass is not protected to
be of $\mathcal O(100~\mathrm{GeV})$ to break the electroweak
symmetry. In SUSY, the loop diagrams, shown in Fig.\ref{canc},
that are quadratically divergent cancel, term by term against the
equivalent diagrams involving superpartners:

\begin{eqnarray*} m^2_h &=& m^2_{h,tree}+c\frac{g^2}{4
\pi^2}M^2_{pl},\,\,\,\,\,\,\,\,\,\, without\,\, SUSY
\nonumber\\
m^2_h &=& m^2_{h,tree}\bigg(1+c'\frac{g^2}{4 \pi^2}\ln\big(
\frac{M_{pl}}{M_W}\big)\bigg)\,\,\,\,\,\,\,\,\,\, with\,\,
SUSY\end{eqnarray*}

If $m_h \sim  O(100)$ GeV, the masses of superpartners should be $
\lesssim O(1)$ TeV. Thus, some of the superpartners will be
detected at the LHC.

\begin{figure}[t]
\begin{center}
\includegraphics[width=7cm,height=4.5cm]{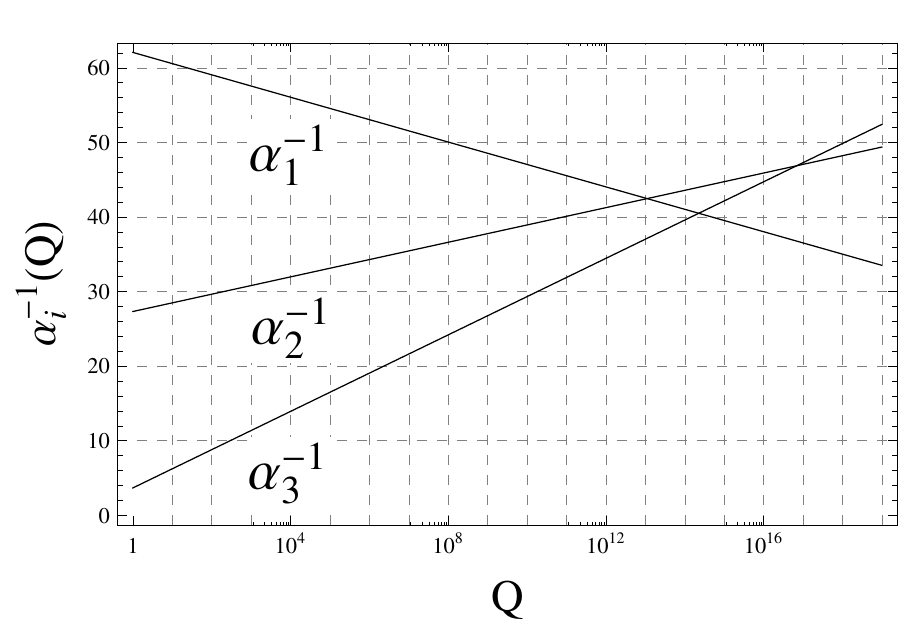}~~~~~~\includegraphics[width=7cm,height=4.5cm]{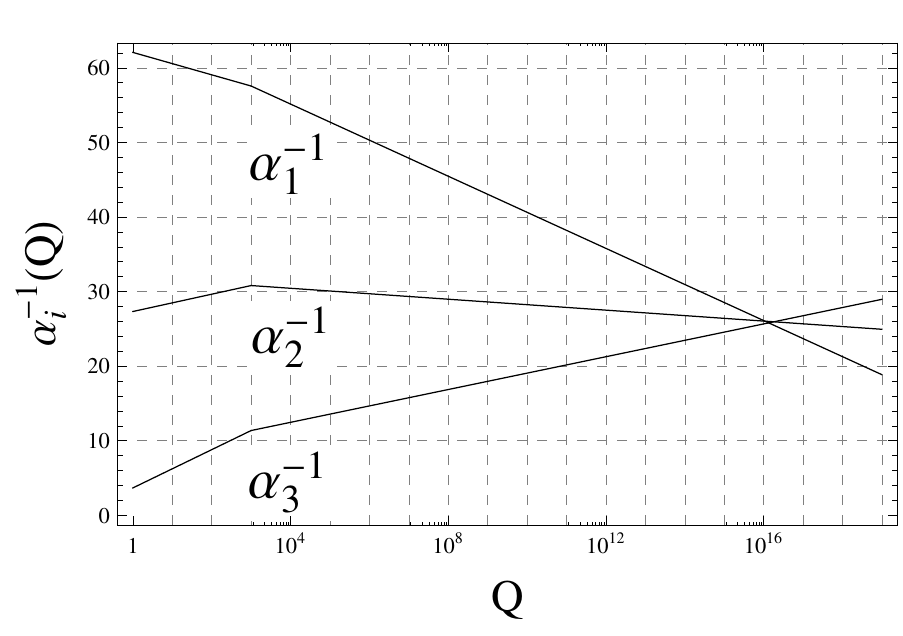}
\caption{Gauge Coupling Unification.} \label{unif}
\end{center}
\end{figure}

Additional support for low scale ($\sim 1 $ TeV) SUSY follows from
gauge coupling unification. Within SM, the gauge coupling
constants describing the strengths of the electroweak force, the
weak and strong nuclear forces do not unify if they run to high
energies using the renormalization group equations of these
coupling constants while within SUSY they do as can be seen in
Fig.\ref{unif}.

As known, the formalism of SM  does not allow proper incorporation
of the gravitational interactions in its gauge group symmetry. On
the other hand, the Poincar\'{e} group corresponds to the basic
symmetries of the special relativity. It turns out that, in order
to unify gravity with the gauge interactions, we need to combine
Poincar\'{e} and internal symmetries \cite{Quevedo:2010ui}. It
should be noted also that, according to the Coleman-Mandula
theorem, the most general symmetry which quantum field theory can
have is a tensor product of the Poincare group and an internal
group \cite{Coleman:1967ad}. In this context, SUSY is an extension
of the spacetime symmetry reflected in the Poincar\'{e} group.
Thus, upon  space-time $(x^\mu) ~ \to$ ~ Superspace $(x^\mu,
\theta^\alpha)$, SUSY is a translation in Superspace given as:
\begin{eqnarray*}
x^\mu &\to&  x'^\mu = x^\mu + \frac{i}{2} \bar{\epsilon}\gamma^\mu \theta \\
\theta &\to&  \theta' = \theta + \epsilon
\end{eqnarray*}

Linking bosonic degrees of freedom to fermionic ones can be
generated in SUSY by an operator $Q$ that carries spin-1/2 acting
as

\begin{equation}
Q\left|\mathrm{Boson}\right>=\left|\mathrm{Fermion}\right>,~~~~~~~~~~Q\left|\mathrm{Fermion}\right>=\left|\mathrm{Boson}\right>.
\end{equation}

The simplest and most useful supersymmetry algebra in four
dimension Minkowski space is is obtained by adding to
Poincar\'e algebra a Majorana spinor charge $Q_{\dot\alpha}%
,\dot\alpha=1,2$, satisfying the property
 \begin{align}
 \left\{  Q_{\alpha},\bar{Q}_{\dot{\alpha}}\right\}   &
 =2\sigma_{\alpha
 \dot{\alpha}}^{\mu}P_{\mu},
 \end{align}
where $P_\mu$ are the generators of translation and $\sigma^\mu$
are the Pauli and unit matrices. Superfied
$\Phi(x,\theta,\bar{\theta})$, as function of the Superspace
coordinates, can be defined as

\begin{eqnarray*}
\Phi(x, \theta, \bar{\theta})&=& \phi(x)+\theta \psi(x) +
\bar{\theta} \bar{\chi} + \theta \theta\ m(x) + \bar{\theta}
\bar{\theta}\ n(x) + \theta \sigma^\mu \bar{\theta}\ v_\mu(x) \\
&+& \theta \theta \bar{\theta}\ \bar{\lambda}(x)+ \bar{\theta}
\bar{\theta}\theta\ \eta(x)+ \theta \theta \bar{\theta}
\bar{\theta}\ d(x),
\end{eqnarray*}

Chiral Superfields, corresponding to a Weyl fermion and a complex
scalar, must satisfy the condition $\bar{D} \Phi=0$ where $D$ is a
derivative operator in the Superspace. As a consequence, chiral
Superfield can be expressed as
\begin{equation}\Phi(x,\theta) = \phi(x) + \sqrt{2}\theta \psi(x) + \theta \theta
F(x)
\end{equation}

where $\phi(x)$, $\psi(x)$ and F(x) denote a complex scalar, a
Weyl fermion and auxiliary fields respectively.  The infinitesimal
SUSY transformation of chiral superfield yields $\Phi \to \Phi +
\delta \Phi$ with $\delta \Phi = i(\xi Q + \bar{\xi} \bar{Q})
\Phi$ and implies
\begin{eqnarray*}
\delta_{\xi} \phi &=& \sqrt{2} \xi \psi, \label{delta-phi}\\
\delta_{\xi} \psi &=& \sqrt{2} \xi F - \sqrt{2} i \sigma^\mu
\bar{\xi} \partial_\mu \phi, \label{delta-psi}\\
\delta_{\xi} F &=& \sqrt{2} i   \psi \sigma^\mu
\bar{\xi}\partial_\mu.\label{delta-F}
\end{eqnarray*}

where $\delta F$ is a total derivative. Thus, if a Lagrangian is
made out of the highest component of a superfield, it is SUSY
invariant. In SUSY, vector Superfields corresponds to a gauge
boson (massless vector) and a Weyl fermion and defined via the
requirement $V=V^+$:

\begin{eqnarray*}
V(x,\theta, \bar{\theta}) &=&\!\!\!\! C(x) + i \theta \chi(x) - i
\bar{\theta} \bar{\chi}(x) + \theta \sigma^\mu \bar{\theta}
v_{\mu} +  \frac{i}{2} \theta \theta \left[M(x) + i N(x)\right]\nonumber \\
&-&\!\!\!\!\frac{i}{2} \bar{\theta} \bar{\theta} \left[M(x) - i
N(x)\right] + \theta \theta \bar{\theta} \left[\bar{\lambda}(x) +
\frac{i}{2}
\bar{\sigma}^\mu \partial_\mu \chi(x) \right]\nonumber\\
&+&\!\!\!\!\bar{\theta} \bar{\theta} \theta \left[\lambda(x) -
\frac{i}{2} \sigma^\mu \partial_\mu \bar{\chi}(x) \right] +
\frac{1}{2} \theta \theta \bar{\theta} \bar{\theta} \left[D(x) -
\frac{1}{2}
\partial_\mu \partial^\mu C(x)\right]\!\!,%
\label{vector}
\end{eqnarray*}
to reduce their number, we introduce a generalization of the usual
concept of gauge transformations of spinor and gauge field to the
case of chiral and vector
superfields:%
$$V \to V + \Lambda + \Lambda^\dag, $$
where $\Lambda$ is a chiral superfield. Under this
transformation, the real vector field $v_\mu$ transforms as %
\begin{equation} v_\mu (x) \to v_\mu + i \partial_\mu \left[\alpha(x)
-\alpha^*(x)\right],\end{equation} leading to

\begin{equation}V(x,\theta,\bar{\theta}) = -\theta \sigma^\mu \bar{\theta} v_\mu +
i \theta^2 \bar{\theta}_{\dot{\alpha}}
\bar{\lambda}^{\dot{\alpha}} - i \bar{\theta}^2 \theta_{\alpha}
\lambda^\alpha + \frac{1}{2} \theta^2 \bar{\theta}^2
D(x)\end{equation} here $D(x)$ is a non-propagating auxiliary
field, it transforms under a SUSY transformation into a total
derivative.

\begin{table}[t]
\caption{MSSM Particle Content.} \centering\begin{tabular}{ll} \br
  Gauge bosons S=1 & Gauginos  S=1/2 \\
  gluon,\,$W^\pm,Z,\gamma$ & gluino,\,$\widetilde W,\widetilde Z,\widetilde \gamma$  \\\mr
  Fermions S=1/2 & Sfermions S=0 \\
   $\left(%
\begin{array}{c}
  u_L \\
  d_L \\
\end{array}%
\right) $\,\, $\left(%
\begin{array}{c}
  \nu^e_L \\
  e_L \\
\end{array}%
\right) $ &  $\left(%
\begin{array}{c}
  \widetilde u_L \\
  \widetilde d_L \\
\end{array}%
\right) $\,\, $\left(%
\begin{array}{c}
 \widetilde \nu^e_L \\
 \widetilde  e_L \\
\end{array}%
\right) $ \\
$u_R,\,d_R,\, e_R$ & $\widetilde u_R,\, \widetilde d_R,\, \widetilde e_R$ \\
\mr
  Higgs & Higgsinos \\
  $\left(%
\begin{array}{c}
  H^+_1 \\
  H^0_1 \\
\end{array}%
\right) $\,\, $\left(%
\begin{array}{c}
  H^0_2 \\
  H^-_2 \\
\end{array}%
\right) $ &  $\left(%
\begin{array}{c}
 \widetilde H^+_1 \\
 \widetilde H^0_1 \\
\end{array}%
\right) $\,\, $\left(%
\begin{array}{c}
 \widetilde H^0_2 \\
 \widetilde H^-_2 \\
\end{array}%
\right) $ \\
  \br
\end{tabular}\label{MSSMpart}
\end{table}

In terms of the superfields components the most general
renormalisable non-Abelian gauge invariant Lagrangian is given as:
 \begin{eqnarray}
 \mathcal{L}&=& \sum_i \left ( \vert D \phi_i \vert^2 + i\psi_i
 \sigma^{\mu} D_{\mu} \psi_i^* +\vert F_i\vert^2 \right)
 -\sum_a \frac{1}{4g_a^2} \left [ (F_{\mu \nu}^a)^2 -i \lambda^a
 \slash \!\!\!\!D \lambda^{a*} -\frac{1}{2}(D^a)^2 \right]
 \nonumber\\
 &+& i \sqrt{2}\sum_{ia} g^a \psi_i T^a \lambda^a \phi^*_i
 +{\rm h.c.}+ \sum_{ij} \frac{1}{2} \frac{\partial^2 W}{\partial \phi_i
 \partial \phi_j} \psi^i \psi^j,
 \end{eqnarray}
with $D_\mu$, $F_{\mu \nu}$, $\lambda$, $\phi$ and $\psi$
represent the gauge covariant derivative, the field strengths, the
gaugino filds, the scalar and fermionic fields, respectively.
$T^a$ and $g^a$ are being the generators and coupling constants of
the corresponding groups. Eliminating the auxiliary fields $F^i$
and $D^a$, through equations of motion, gives rise to the scalar
potential of the form
  \begin{equation}
  V_{\rm SUSY} = \frac{1}{2} \vert D^a \vert^2 + \vert F^i\vert^2,
  \end{equation}
with $F^i = \partial W/\partial \phi_i$ and $D^a=g^a \sum_i
\phi_i^* T^a \phi_i$.

Up to date, we did not observe squarks and selectrons. If
Supersymmetry is an exact symmetry then all particles in the same
supersymmrtic multiplet would have the same mass. This indicates
that SUSY must be broken symmetry or else SUSY particles should
have been observed with same mass as SM-partners. On the other
hand, the cancellation of quadratic divergences requires SUSY
partners not to be heavier than $\sim$ TeV. Several ways have been
discussed in the literature to break SUSY. From the definition of
the SUSY algebra:
\begin{equation*}
H= \frac{1}{4} ( \bar{Q}_1 Q_1 +Q_1 \bar{Q}_1 + \bar{Q}_2 Q_2 +Q_2
\bar{Q}_2 ) \geq 0,
\end{equation*}

If the vacuum is supersymmetric the $E_{\rm vac}= \langle 0 \vert
H \vert 0 \rangle = 0$ and if SUSY is broken then $ E_{\rm vac} >
0$. Hence SUSY is broken if $ \langle 0 \vert F_i \vert 0 \rangle
\not = 0$ or $\langle 0 \vert D \vert 0 \rangle \not = 0$. One may
introduce terms in the Lagrangian which break SUSY softly i.e.
these terms do not lead to quadratic divergences. The general
structure for the SUSY breaking includes three sectors:

i) Observable sector: which comprises all the ordinary particle
and their SUSY particles,

ii)  Hidden sector: where the breaking of SUSY occurs,

iii) The messengers of the SUSY breaking from hidden to observable
sector.

The soft SUSY breaking terms are: masses for the scalars, masses
for the gauginos, cubic couplings for scalars.

The Minimal Supersymmetric Standard Model (MSSM) is a
straightforward supersymmetrization of the SM with minimal number
of new parameters. The particle content of the MSSM, see Table
\ref{MSSMpart}, consists of two Higgs doublet SM, scalar SUSY
partners and fermionic SUSY partners. The superpotential of the
MSSM  is given by
\begin{equation}
W_{\rm MSSM}=Y_{u}^{ij}Q_{L_{i}}U_{L_{j}}^{C}H_{u}+Y_{d}^{ij}Q_{L_{i}}D_{L_{j}%
}^{C}H_{d}+Y_{e}^{ij}L_{L_{i}}E_{L_{j}}^{C}H_{d}+\mu H_{d}H_{u},
\end{equation}
the indices $i$ and $j$ refer to quark and lepton families. The
parameters $Y_{u}^{ij},$ $Y_{d}^{ij}$ and $Y_{e}^{ij}$ correspond
to the Yukawa couplings present in the SM, which are non-diagonal
$3\times3$ matrices in flavor space. The $\mu$ parameter has mass
dimension.  It should be noted that, a new symmetry,  R-symmetry
has been introduced to forbid $B-L$ violating interactions in the
superpotential(i.e, no proton decay). In the MSSM, universal soft
SUSY breaking terms includes Universal scalar mass $m_0$,
Universal gauging mass $M_{1/2}$ and Universal trilinear coupling
$A_0$. These terms induce about 100 free parameters which reduce
the predictivity of the MSSM. However, at a specific high scale
models, these parameters can be reduced through the relations
among them as in the constrained MSSM (mSUGRA).

\begin{figure}[t]
\begin{center}
\vskip -0.5cm
\includegraphics[scale=0.4]{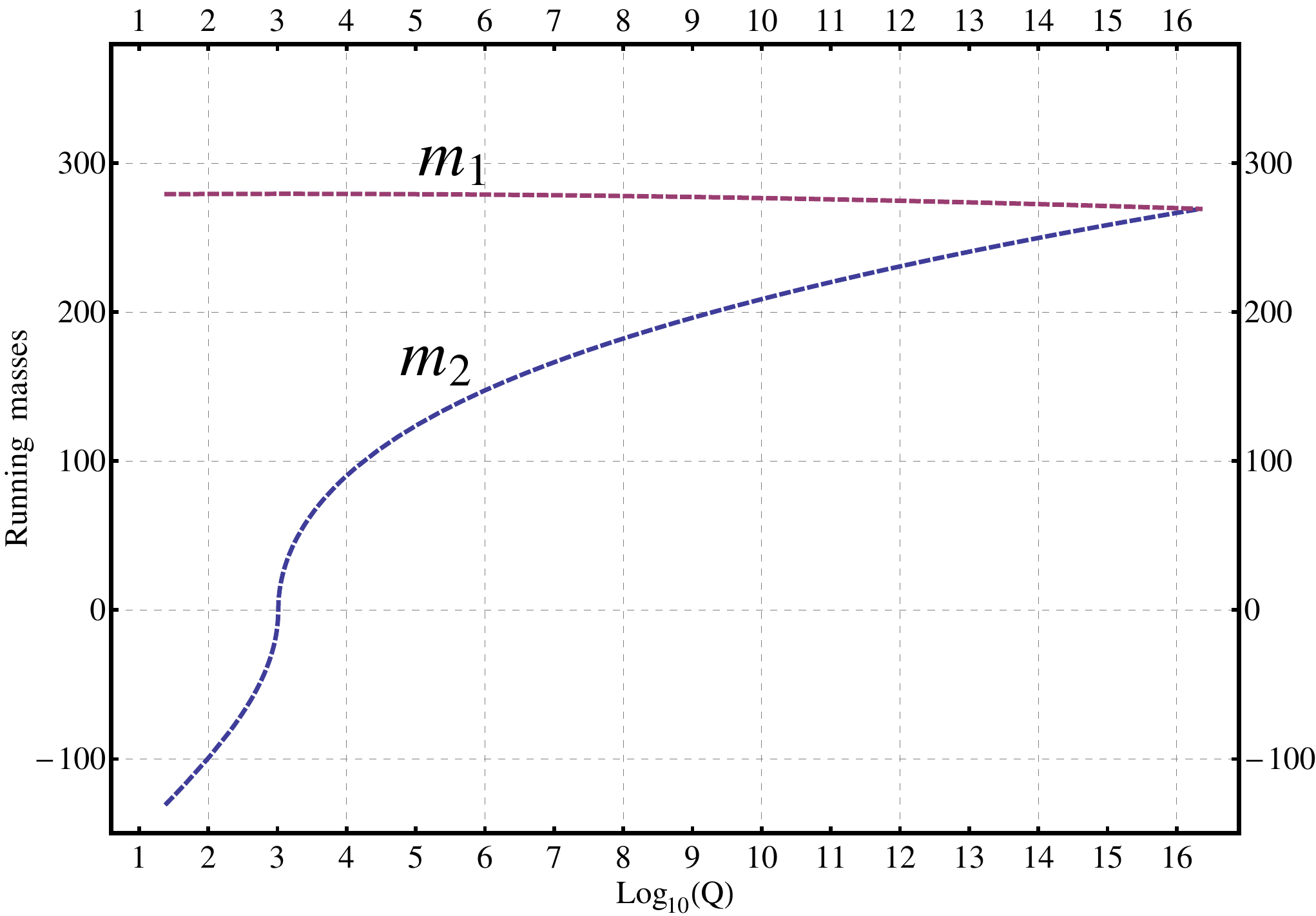}%
\caption{RGE-MSSM.\label{b2smm1}}
\end{center}
\end{figure}

SUSY can provide a natural mechanism for understanding Higgs
physics and electroweak symmetry breaking (EWSB). In the MSSM,
$H_u$ couples to a $t$ (s)quark with a large Yukawa coupling,
unlike $H_d$, which couples to a $b$ (s)quark and a $\tau$
(s)lepton. The Yukawa coupling gives a negative contribution to
the squared masses $m^2_{H_{u,d}}$. The running from $M_X$ down to
the EW scale, shown in Fig.(\ref{b2smm1}), reduces the squared
Higgs masses until, eventually, conditions satisfied and the gauge
symmetry is broken. This is an appealing feature in SUSY models
that generally explains the mechanism of the EWSB dynamically.

 The Higgs sector of the MSSM consists of two complex Higgs
doublets, $H_u$ and $H_d$. After  EWSB, three of the eight degrees
of freedom are eaten by  $W^{\pm}$ and $Z$. The five physical
degrees of freedom that remain form a neutral pseudoscalar (or
CP-odd) Higgs boson $A$, two neutral scalars (or CP-even) $h$, and
$H$ and a charged Higgs boson pair (with mixed CP quantum numbers)
$H^{\pm}$. The mass of the lightest CP-even (SM-like) Higgs, at the one-loop level, is given by  %
$ m_h^2 \leq M_Z^2 + \frac{3 g^2}{16\pi^2 M_W^2}
\frac{m_t^4}{\sin^2\beta} \log\left(\frac{m_{\tilde{t}_1}^2
m_{\tilde{t}_2}^2}{m_t^4}\right)$. MSSM predicts an upper bound
for the Higgs mass: $m_h \lesssim 130$~GeV, which was consistent
with the measured value of Higgs mass (of order 125~GeV) at the
LHC. This mass of lightest Higgs boson implies that the SUSY
particles are quite heavy. This may justify the negative searches
for SUSY at the LHC-run I. In the MSSM, the gluino mass
$m_{\tilde{g}} \simeq 2.5~m_{1/2}$. Thus, due to the LHC
constraint shown in Fig.(\ref{b2smm1}), we conclude that
$m_{\tilde{g}}\geq 1.5$~TeV for a value $m_{1/2}\geq 600$~GeV.

\section{SUSY Dark Matter }
\vskip 0.2cm

One of the longstanding problems that maybe considered as a hint
to the necessity of existence of physics beyond SM is dark matter
in the Universe\cite{Zwicky:1933gu}. The presence of a such
matter, which is different from the familiar baryonic matter, is
supported by astrophysical observations and cosmological
considerations.  The relic abundance of dark matter is given by
~\cite{Ade:2015xua}
\begin{equation} \Omega_{\rm DM} h^2 = 0.1188 \pm 0.0010 \, ,
\label{eq:omegah}
  \end{equation}
here $h$ denotes the reduced Hubble constant and accounts for
nearly $25\%$ of the total energy of the Universe. So far, it is
believed that DM can be accounted for by a  particle that is
either stable, at least on cosmological scales, or has a lifetime
much larger than that of the universe. This requirement can be
achieved by an appropriate symmetry imposed on the model.
Candidates of such particle must have attractive gravitational
interactions and their other interactions with the SM states
should be very suppressed which can be fulfilled  by electrically
and color neutral particles. This can be understood as up to date
there is no evidence that DM has any other interaction except
gravity. Further more, DM candidates have to be non-relativistic,
i.e. cold,  at the time of matter-radiation equality in the
Universe. The possibility of hot dark matter is ruled out by
several observations such as gravitational growth of small-scale
structure, formation of stars, galaxies, and clusters of galaxies
so early and the weak lensing signals we see and  the pattern of
fluctuations in the cosmic microwave background. In the SM,
neutrinos only can be a candidate for DM. However, neutrinos are
too light to account for the dark matter that must be present in
our Universe. So, we need to look for candidates of DM in beyond
SM physics. In the following we explore these possibilities in
Supersymmtry as an extension of the SM.

In the SM, the mixing of the gauge bosons  ${W^i}$ and $ B^0 $,
after electroweak symmetry breaking, leads to the physical (mass)
states $\gamma ,Z^0 $ and $W^{\pm}$. Similarly, in the MSSM, the
mixing among SUSY partners of the SM fields results in the
neutralinos as new fermionic mass state. Particularly, the neutral
gauginos $(\tilde{B}, \tilde{W}^0)$ and the neutral higgsinos
 $(\tilde{H}^0_{u,d})$ mixing form
 four neutral mass-eigenstates called neutralinos.
 If we define a gauge-eigenstate
basis \begin{equation}
{\psi}^0=\left( \ba{c} {\tilde{B}} \\ {\tilde{W}}^0\\ {\tilde{H}}^0_{\rm d} \\
{\tilde{H}}^0_{\rm u} \ea \right),
\end{equation}
one cane write
\begin{equation}
- \frac{1}{2} {\psi}^{0 {\rm T}}  {\small\bf m_{\tilde{\chi}^0}}
{\psi}^0 + {\rm h.c.}
\end{equation} with

\begin{equation} {\small\bf m_{\tilde{\chi}^0} = \left(
\begin{array}{cccc}
M_1 &0 &-\frac{1}{2} g_1 v_d  &\frac{1}{2} g_1 v_u  \\
0 &M_2 &\frac{1}{2} g_2 v_d  &-\frac{1}{2} g_2 v_u  \\
-\frac{1}{2} g_1 v_d  &\frac{1}{2} g_2 v_d  &0 &- \mu  \\
\frac{1}{2} g_1 v_u  &-\frac{1}{2} g_2 v_u  &- \mu  &0
\end{array}
\right),} \end{equation}

here $v_{d} = \langle H_{d}\rangle$ and $v_{u} = \langle
H_{u}\rangle$. The parameters $M_1$, $M_2$ and $\mu$ can have
arbitrary phases. However, one can redefine the phases of
${\tilde{B}}$ and ${\tilde{W}}^0$ to make both $M_1$ and $M_2$
real and positive. Usually, $\mu$ is  taken to be be real to avoid
unacceptably large {\bf CP}-violating effects including EDM for
both the electron and the neutron. Hence, the neutralino mass
matrix ${\small\bf m_{\tilde{\chi}^0}}$ is real and symmetric and
therefore, it can be diagonalized analytically
\cite{ElKheishen:1992yv} by a single $4\times4$ real matrix N such
that \be N^* m_{\tilde{\chi}^0}
N^{-1}=dig(m_{\tilde{\chi}^0_1},m_{\tilde{\chi}^0_2},
m_{\tilde{\chi}^0_3},m_{\tilde{\chi}^0_4}) \ee  The resulting four
neutral mass eigenstates are called neutralinos, with the
convention that the masses are ordered as $m_{{\tilde{\chi}}^0_1}
< m_{{\tilde{\chi}}^0_2} < m_{{\tilde{\chi}}^0_3}
<m_{{\tilde{\chi}}^0_4}$. The physical Majorana neutralino (mass
eigenstates) can be written as \cite{Cho:1996we} \begin{equation}
\tilde{\chi}^0_M=N\left(\begin{array}{c}
  \widetilde{B} \\
  \widetilde{W^0} \\
  \widetilde{H^0_d} \\
   \widetilde{H^0_u} \\
\end{array}\right)+N^*C \left(\begin{array}{c}
 { \overline{\widetilde{B}}}^{\,T} \\
{  \overline{\widetilde{W^0}}}^{\,T} \\
  {\overline{\widetilde{H^0_d}}}^{\,T} \\
  {\overline{\widetilde{H^0_u}}}^{\,T} \\
\end{array}\right)
\end{equation}

\begin{figure}[t]
\begin{center}
\vskip -0.5cm
\includegraphics[scale=0.6]{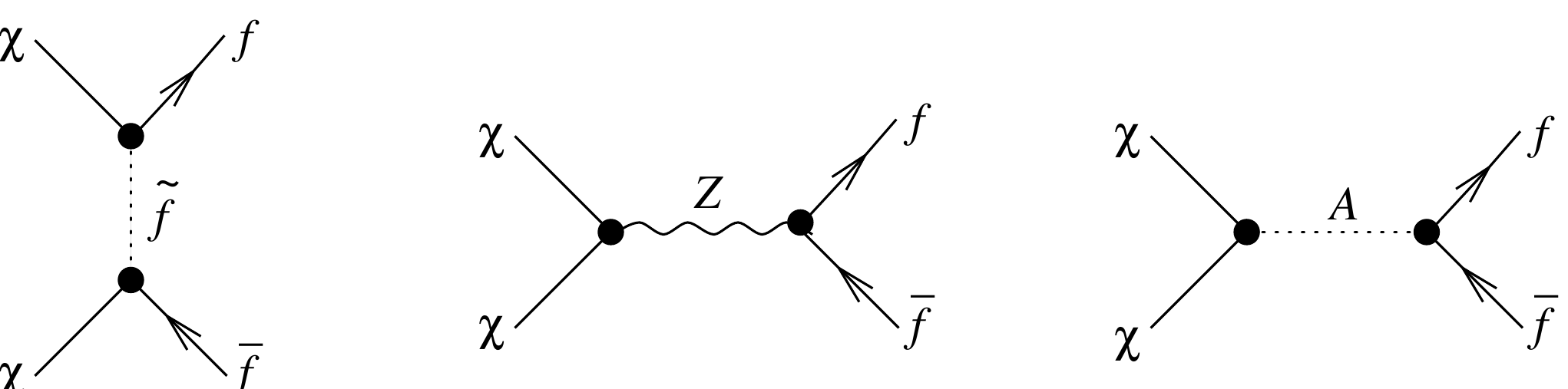}%
\caption{$\tilde{\chi}^0_1$ annihilation into fermions through
sfermions, $Z$ gauge boson and Higgs states
\cite{Jungman:1995df}.\label{b2smm}}
\end{center}
\vskip -0.5cm
\end{figure}

In the limit $\vert\mu\vert\rightarrow\infty$, $\tilde{\chi}^0_1$
corresponds to a pure bino with mass $m_{\tilde{\chi}^0_1}\simeq
M_1$, $\tilde{\chi}^0_2$ corresponds to pure wino with mass
$m_{\tilde{\chi}^0_2}\simeq M_2$ while $\tilde{\chi}^0_3$ and
$\tilde{\chi}^0_4$ are pure higgsinos with masses
$m_{\tilde{\chi}^0_3}\simeq m_{\tilde{\chi}^0_4}\simeq \vert
\mu\vert$. In mSUGRA, we have \cite{Martin:1997ns} \be
M_1\approx\frac{5}{3}\tan^2\theta_W M_2\approx 0.5M_2 \ee
 which indicates that  \be
 M_1<M_2\ll\vert\mu\vert
 \ee  and therefore, $\tilde{\chi}^0_1$ is the lightest
neutralino and usually it is the lightest SUSY particle. Since, it
is also electrically neutral and has no color charge, it is an
attractive candidate for non-baryonic dark matter
\cite{Ellis:1983ew} if being stable. In fact $R$-parity symmetry
can ensure that $\tilde{\chi}^0_1$ is stable. The conservation of
the R- parity symmetry  implies that
 SUSY particles can only be produced (destroyed) in pairs
form (into) SM particles.  Moreover, heavy unstable SUSY particle
will decay in a chain until the lightest SUSY particle (`LSP'),
$\tilde{\chi}^0_1$, is produced. Thus, the stability of
$\tilde{\chi}^0_1$ is guaranteed by the R-parity symmetry. As a
result, within SUSY models conserving R-parity symmetry, the LSP
is a good candidate of DM.

We turn now to discuss the calculations of the cosmological
neutralino relic abundance. Generally, all SUSU  scalar particles
can contribute to $\Omega_{\tilde{\chi}^0_1} h^2$ as they decay
until finally LSPs are produced, and all the (co)annihilation
processes must be considered. However, the most important
contributions to the neutralino relic density come from the LSP.
In this case, the most important final states into which the
neutralino can annihilate include the two-body final states which
occur at tree level. In particular, these states are, SM
fermion-antifermion pairs (see Fig.\ref{b2smm}), SM massive gauge
bosons $W^+ W^-,Z^0 Z^0$ and $Z^0 h$. Other states includes one SM
massive gauge boson in addition to one SUSY Higgs boson, pair of
charged SUSY Higgs and a combination of SM Higgs with one of the
neutral SUSY Higgs bosons. For detailed calculations and
discussion, we refer to Ref.\cite{Jungman:1995df}.

The MSSM parameter space consists of 100 free parameters. However,
they are highly constrained by flavor and CP-violating
observables. In fact, with the discovery of the Higgs boson at
LHC, the lack of positive signals from direct searches at the LHC,
and null results from direct detection experiments, the MSSM turns
to be almost ruled out.

\begin{figure}[t]
\begin{center}
\vskip -0.5cm
\includegraphics[scale=0.7]{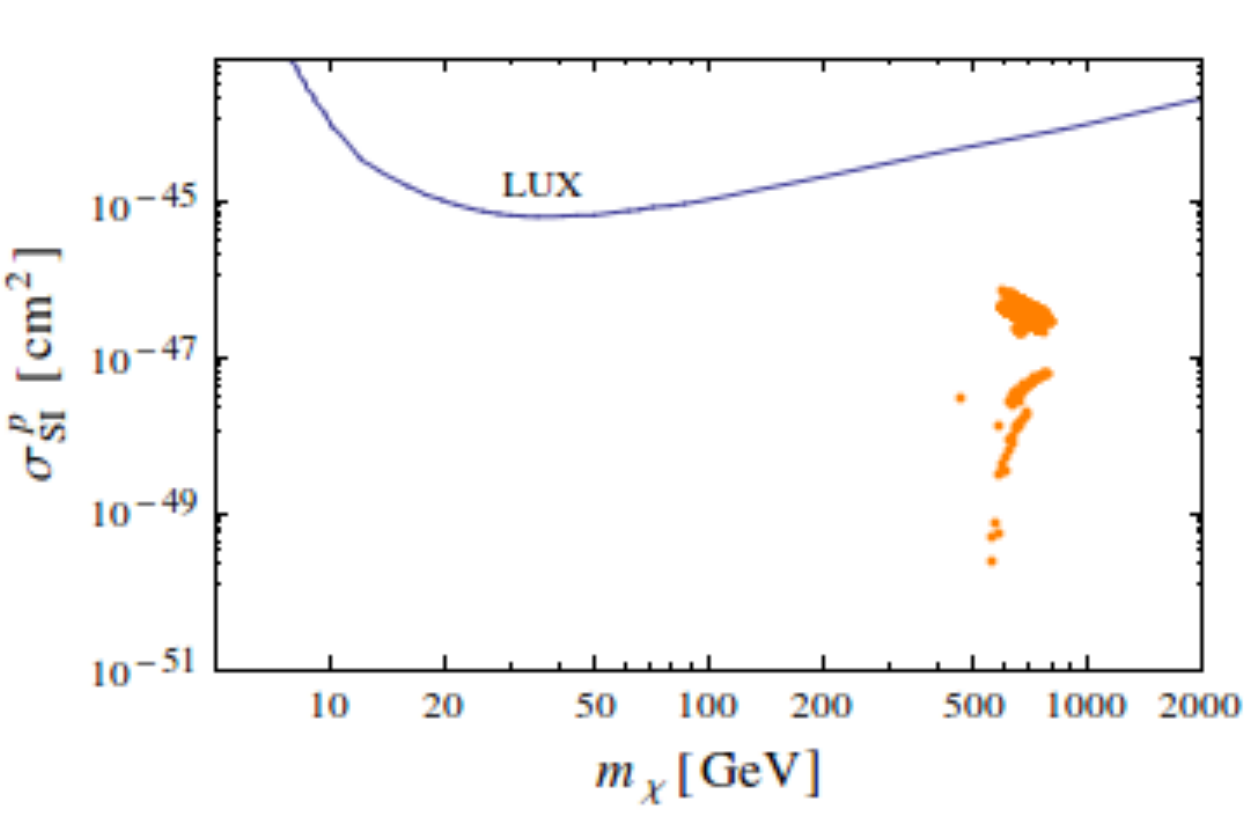}
\caption{Spin-independent scattering cross section of the LSP with
a proton versus the mass of the LSP within the region allowed by
all constraints (from the LHC and relic abundance.\label{b2f}}
\end{center}
\vskip -0.5cm
\end{figure}

The cross section for elastic scattering of a neutralinos off
ordinary mater can determine the detection rate in the
direct-detection experiments. Basically, this scattering depends
on the strength of the neutralino quark interaction, the
distribution of quarks inside the nucleon and the distribution of
nucleons inside the nucleus []. The measured nuclear recoil energy
resulting from this elastic scattering, in direct-detection
experiments, can serve as a direct search of neutralinos as DM
particles.  At tree-level, the elastic scattering of the
neutralino ($\tilde{\chi}^0_1$) off nucleus, mediated by Squarks
($\tilde q$), Higgse ($H$) and $Z$ exchange. The Spin-independent
scattering cross section of the LSP with a proton versus the mass
of the LSP within the region allowed by all constraints (from the
LHC and relic abundance) is shown in Fig.(\ref{b2f}). Clearly, we
need to go beyond MSSM for a possible candidate of DM.

\section{Non-Minimal Supersymmetric Standard Model}
\vskip 0.2cm

The solid experimental evidence for neutrino oscillations,
pointing towards non-vanishing neutrino masses, is one of the few
firm hints for physics beyond the SM. In the SM, the global
($B-L$) symmetry, where $B$ and $L$ stand for baryon and lepton
numbers respectively, is conserved. Extending the MSSM by gauging
this symmetry, based on the group $SU(3)_c \times SU(2)_L \times
U(1)_Y \times U(1)_{B-L}$, resulting in the so-called B-L
Supersymmetric(SUSY) model, BLSSM, and can be considered as the
minimal extension of MSSM with significantly enriched particle
content. In fact, gauging $(B-L)$ symmetry requires adding three
SM singlet fields to cancel the triangle anomaly diagrams. These
singlet fields may be identified as the right handed (RH)
neutrinos. Within this model, the light Left-Handed (LH) neutrino
masses can be generated through either a low (TeV) scale Type-I
see-saw mechanism or inverse seesaw mechanism. Another feature of
the model is the possibility to spontaneously break the ($B-L$)
symmetry with new Higgses, usually known as bileptons. In turns,
the new $Z'$  gauge boson associated with this group, will acquire
a mass. As in the case of MSSM, their will be superpartners of the
new particles in the model namely, the RH sneutrinos, the
superpartners of bileptinos and the superpartner of the new $B'$
boson, the BLino. Besides, the BLSSM have the same features of the
MSSM such as gauge coupling unification, solution of the hierarchy
problem and others with several new DM candidates as we will
discuss in the following.

Regarding type-I seesaw mechanism, right-handed neutrinos can
acquire Majorana masses at the scale of B-L symmetry breaking. On
the other hand, in the inverse seesaw mechanism, B-L gauge
symmetry does not allow these Majorana masses.  As a consequence,
another pair of SM gauge singlet fermions with masses of
~${\mathcal O}(1)$ keV has to be introduced. One of the two
singlet fermions plays a rule in generating the light neutrino
masses through its couplings to the right handed neutrino while
the other singlet is completely decoupled and can interacts only
through the B-L gauge boson and consequently play the role of warm
dark mater as we will see below. In the SUSY B-L Model with
Inverse Seesaw (BLSSM-IS), the most general superpotential of the
model can be given as
\begin{eqnarray*}
\hspace{-0.25cm}W \!=\!  - {\mu_{\eta}} \hat{\chi}_1\hat{\chi}_2
\!+\!\mu\hat{H}_u\hat{H}_d \!+\!\mu_S\hat{S}_2\,\hat{S}_2 \!-\!
Y_d \hat{d}\hat{q}\hat{H}_d \!-\! Y_e \hat{e}\hat{l}\hat{H}_d
\!+\!Y_u \hat{u}\hat{q} \hat{H}_u \!+\!Y_s \hat{\nu}
\hat{\chi}_1\hat{S}_2 \!+\!Y_\nu\,\hat{\nu}\hat{l}\hat{H}_u.
\end{eqnarray*}
where $\hat \chi_{1,2}$ are SM singlet chiral superfields with
$B-L$ charges $+1$ and $-1$, respectively. The VEVs of the scalar
components of these superfields breaks $U(1)_{B-L}$ spontaneously.
In the superpotential $\hat{\nu}$ represents three chiral singlet
superfields with $U(1)_{B-L}$ charge $= -1$. The three chiral SM
singlet superfields $\hat S_{1,2}$ with $B - L$ charge $= +2,-2$
are considered to implement the inverse seesaw mechanism and a
${Z}_2$ symmetry is assumed to forbid the interactions between
$S_1$ and other fields.

 The SUSY soft breaking Lagrangian is given by
\begin{eqnarray*} -\mathcal{L}_{\tiny\textnormal{soft}} \!\!&\!
\!=\! \!&\! \! m_0^2\Big[|\tilde{q}|^2 + |\tilde{u}|^2 +
|\tilde{d}|^2 + |\tilde{l}|^2 + |\tilde{e}_R^*|^2 +
|\tilde{\nu}_R^*|^2 + |\tilde{S}_1|^2 + |\tilde{S}_2|^2 + |H_d|^2
+ |H_u|^2 +|\chi_1|^2 \nonumber\\ &+&  |\chi_2|^2\Big]+ \left[
Y_u^A \tilde{q} H_u \tilde{u}^*_R + Y_d^A \tilde{q} H_d
\tilde{d}^*_R + Y_e^A \tilde{l} H_d \tilde{e}^*_R + Y_{\nu}^A
\tilde{l} H_u \tilde{\nu}^*_R + Y_s^A \tilde{\nu}^*_R \chi_1
\tilde{S}_2 \right]\nonumber\\ &+& \left[ B ( \mu H_1 H_2 + \mu'
\chi_1 \chi_2 ) + h.c. \right] + \frac{1}{2} M_{1/2} \left[
\tilde{g}^a \tilde{g}^a + \tilde{W}^a \tilde{W}^a + \tilde{B}
\tilde{B} + \tilde{B}' \tilde{B}' + h.c. \right],\nonumber
\end{eqnarray*}
where the trilinear terms are defined as $(Y_f^A)_{ij} = ( Y_f A
)_{ij}$ with $f=u,d,e,\nu,S$.

After the $B-L$ and  EW symmetry breaking, the neutrinos mix with
the fermionic singlet fields. In the flavor basis,
the Lagrangian of neutrino masses, can be expressed as %
\be%
{\cal L}_m^{\nu} =\mu_s \bar{S}^c_2 S_2 +(m_D \bar{\nu}_L \nu_R + M_R \bar{\nu}^c_R S_2 +h.c.) ,%
\ee%
where $m_D=\frac{1}{\sqrt{2}}Y_\nu v$ and $ M_R = \frac{1}{\sqrt
2}Y_{s} v' $. Defining $\psi=(\nu_L^c ,\nu_R, S_2)$,  the neutrino
mass matrix can be written as ${\cal M}_{\nu} \bar{\psi}^c \psi$
with ${\cal M}_{\nu}$ is given by, %

\be  {\cal M}_{\nu}=
\left(%
\begin{array}{ccc}
  0 & m_D & 0\\
  m^T_D & 0 & M_R \\
  0 & M^T_R & \mu_s\\
\end{array}%
\right), \label{nmas}\ee The diagonalization of the mass matrix in
Eq.(\ref{nmas}) leads \cite{Khalil:2010iu}  to the following light
and heavy neutrino masses under the consideration $m_R, \mu_s \ll
m_D, M_R$ \cite{law}

\begin{eqnarray*}%
m_{\nu_l} &=& m_D M_R^{-1} \mu_s (M_R^T)^{-1} m_D^T,\label{mnul}\\
m_{\nu_H}&=& m_{\nu_{H'}} = \sqrt{M_R^2 + m_D^2}. %
\end{eqnarray*} %
light neutrino masses ($\sim$  eV) can be obtained, with a TeV
scale $M_R$, if $\mu_s \ll M_R$  and $Y_{\nu} \sim {\cal O}(1)$.
With a large $Y_\nu$ heavy neutrinos can be probed at the LHC. The
second SM singlet fermion, $S_1$, remains light with mass: ${
m_{S_1} = \mu_s \simeq {\cal O}(1)~ {\rm keV}}$. It is a kind of
sterile neutrino that has no mixing with active neutrino and thus
$S_1$ can therefore be a good candidate for warm dark matter.
$S_1$ can only interact with the B-L gauge boson, $Z'$. It
annihilates through one channel only, into two light neutrinos
mediated by $Z'$. Based on the study in
Ref.\cite{El-Zant:2013nta}, the over-abundance of thermally
produced warm dark matter  can be reduced  to an acceptable range
in the presence of a moduli field decaying into radiation.
Moreover, the warm dark matter candidate can be produced directly
from the decay of the moduli field during reheating. However, as
shown in Ref.\cite{El-Zant:2013nta},  obtaining the right amount
of relic abundance, while keeping the reheat temperature high
enough as to be consistent with Big Bang nucleosynthesis bounds,
sets constraints on the branching ratio for the decay of the
moduli field into dark matter.

  In BLSSM-IS, the one-loop radiative correction of Higgs mass, the
lightest Higgs boson when the latter is Standard Model (SM)-like,
receives significant contribution from right-handed (s)neutrinos
(similar to (s)top effect in MSSM).  These corrections were
estimated in Ref.\cite{Elsayed:2011de} to be be as large as ${\cal
O}(100)$ GeV. This enhancement greatly reconciles theory and
experiment, by alleviating the so-called `little hierarchy
problem' of the minimal SUSY realisation. All three generations of
the (s)neutrino sector may lead to important effects since the
neutrino Yukawa couplings are generally not hierarchical. For
$M_A\gg M_Z$ and $\cos2\beta\simeq 1$, one finds that

\be  m_h^2 \simeq M_Z^2+\delta_t^2+\delta_\nu^2. \ee

 If $\tilde{m}\simeq{\cal O}(1)$ TeV, $Y_\nu\simeq{\cal
O}(1)$ and $M_N\simeq{\cal O}(500)$ GeV, one finds that
$\delta_\nu^2\simeq{\cal O}(100~{\rm GeV})^2$, thus the Higgs mass
is of order $\sqrt{(90)^2+{\cal O}(100)^2+{\cal O}(100)^2}$ GeV
$\simeq 170$ GeV.

The BLSSM-IS has more possibilities for candidates of DM compared
to the MSSM. These includes, the lightest $B-L$ neutralino
($\tilde{B'}$, $\tilde{\eta}_{2}$)-like, referred as
$\tilde{\chi}_1$, and lightest right-handed sneutrino. In the
following, we investigate these possibilities.

\begin{figure}[t]
\begin{center}
\vskip -0.5cm
\includegraphics[width=5cm,height=3.cm]{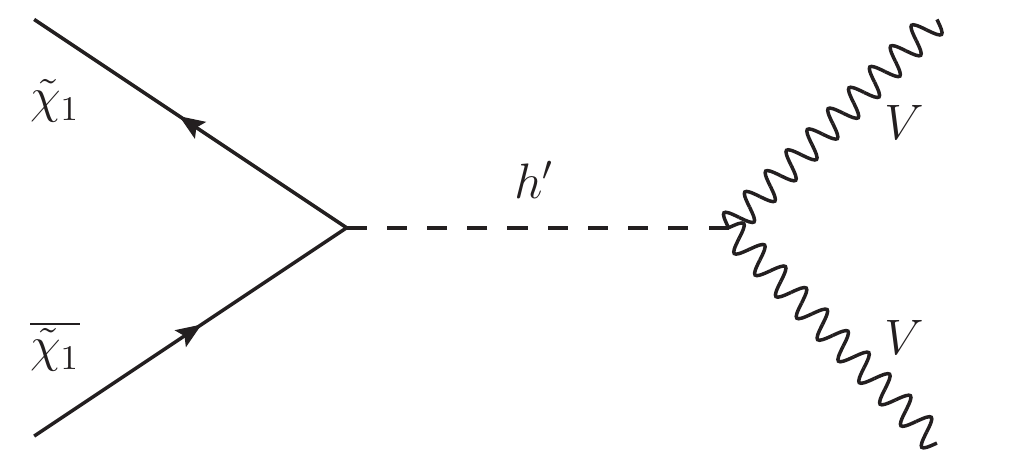}~~~~~~~~~~\includegraphics[width=5cm,height=3.cm]{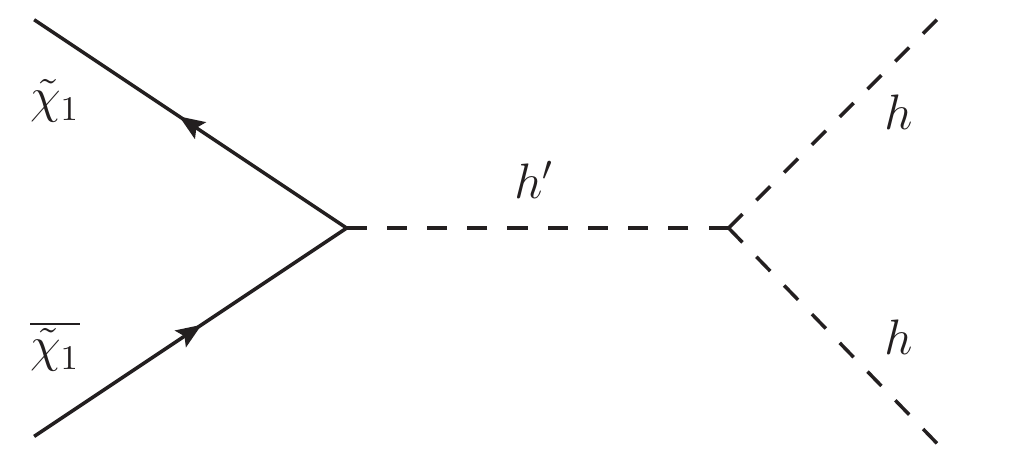}
\caption{Feynman diagrams of the dominant annihilation channels of
the $B-L$ lightest neutralino $\tilde{\chi}_1$ into the SM vector
bosons $(V=W,Z)$ and the SM-like Higgs $h$ mediated by the
lightest $B-L$ CP-even Higgs.} \label{annihilationchi}
\end{center}
\vskip -0.5cm
\end{figure}

In the BLSSM-IS, the neutralinos $\tilde{\chi}^0_i $ (with $i$
running from $1$ to $7$) are the mass eigen states resulting from
the superpositions of three fermionic partners of neutral gauge
bosons, $\tilde{B}$ (bino), $\tilde{W}^3$ (wino) and $\tilde{B'}$
($B'$ino), in addition to the fermionic partners of neutral MSSM
Higgs ($\tilde{H}_1^0$, and $\tilde{H}_2^0 $.) and the fermionic
partners of $B-L$ scalar bosons ($\tilde{\eta}_1$, and
$\tilde{\eta}_2 $). The  neutralino mass matrix ${\cal M}_7$ can
be expressed as \cite{Abdallah:2017gde} \be {\cal M}_7 \equiv
\left(\begin{array}{cc}
{\cal M}_4 & {\cal O}\\
 {\cal O}^T &  {\cal M}_3\\
\end{array}\right),
\ee%
here  ${\cal M}_4$ is the MSSM neutralino mass matrix
\cite{Haber:1984rc,Gunion:1984yn,ElKheishen:1992yv,Guchait:1991ia},
while ${\cal M}_{3}$ is an additional $B-L$ neutralino mass matrix
and ${\cal O}$ is off-diagonal matrix with
\be%
{\cal M}_3 = \left(\begin{array}{ccc}
M_{B'} & -g_{_{BL}}v'_1  & g_{_{BL}}v'_2 \\
-g_{_{BL}}v'_1 & 0 & -\mu'  \\
g_{_{BL}}v'_2 & -\mu' & 0\\
\end{array}\right),~~~~~~{\cal O} = \left(\begin{array}{ccc}
\frac{1}{2}M_{BB'} &~~~0~~~& 0 \\
0 & 0 & 0  \\
-\frac{1}{2}\tilde{g}v_1 &~~~0~~~& 0\\
\frac{1}{2}\tilde{g}v_2&~~~0~~~&0\\
\end{array}\right),\label{mass-matrix.1}
\ee here $M_{B'}$ denotes $B'$ino mass and $M_{BB'}$ represents
the mass mixing term of $\tilde{B}$ and $\tilde{B'}$. At the GUT
scale, $M_{B'}=m_{1/2}$ and $M_{BB'}=0$. Thus, when $\tilde{g}=0$,
the matrix $\cal O$ turns to a zero matrix and the real matrix
${\cal M}_{7}$ can be diagonalized with a symmetric mixing matrix
$V$ such as
\be V{\cal M}_7V^{T}={\rm
diag}(m_{\tilde\chi^0_i}),~~i=1,\dots,7.\label{general} \ee

The LSP $\tilde\chi_1$, in this case, is given by   \be
\tilde\chi_1=V_{11}{\tilde B}+V_{12}{\tilde W}^3+V_{13}{\tilde
H}^0_1+V_{14}{\tilde H}^0_2+V_{15}{\tilde B'}+V_{16}{\tilde
\eta_1}+V_{17}{\tilde \eta_2}. \ee

Clearly, the LSP can be either pure $B'$ino ($\tilde{B'}$) if
$V_{15}\sim1$ and $V_{1i}\sim0$ for $i\neq5$, or pure $B-L$
higgsino $\tilde\eta_{1(2)}$ if $V_{16(7)}\sim1$ and all the other
coefficients are close to zero value. It should be noted that, the
off-diagonal elements $({\cal M}_3)_{12,13}$ and  $({\cal
M}_3)_{21,31}$ are not suppressed. Consequently, unless $\mu'$ is
very large, the lightest $B-L$ neutralino is a mixed state between
$B'$ino and $\tilde{\eta}_{1,2}$.

\begin{figure}[t]
\begin{center}
\vskip -0.5cm
\includegraphics[width=7cm,height=5.5cm]{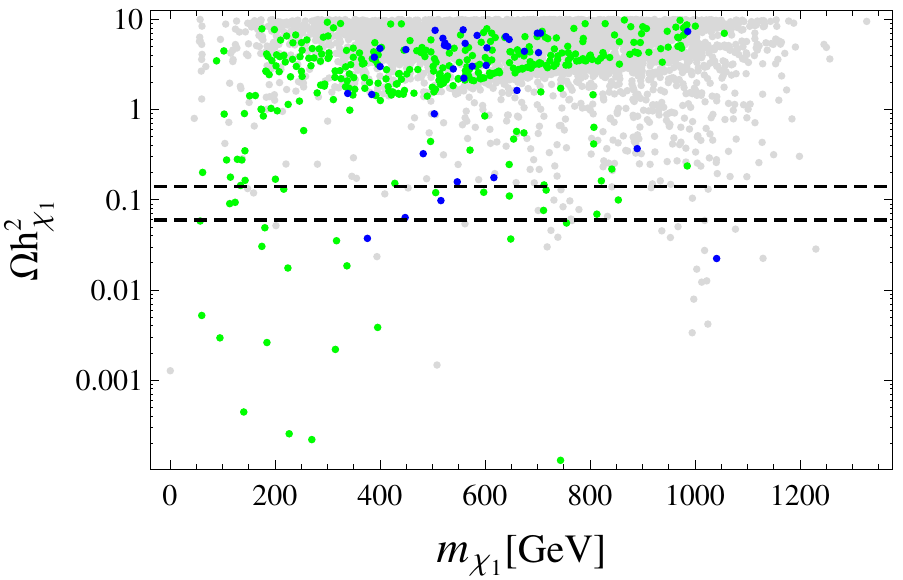}
\caption{The thermal relic abundance of $B-L$ neutralinos,
$\tilde{B'}$-like (green points) and $\tilde{\eta}_2$-like (blue
points), LSP as a function of their masses. Horizontal lines
correspond to the Planck limits on DM abundance. The gray points
indicate to the excluded points by the LHC and LEP constraints.}
\label{MchiOGh2ALL1}
\end{center}
\vskip -0.5cm
\end{figure}

We consider now $\tilde{\chi}_1$ as a DM candidate under the
assumption that $\tilde{\chi}_1$ was in thermal equilibrium with
the SM particles in the early universe where the decoupling
occurred when $\tilde{\chi}_1$ was non-relativistic. In this case,
the dominant annihilation channels of $\tilde{\chi}_1$ are those
channels with final states $W W, ~ZZ, ~ h h $ and are mediated by
the lightest $B-L$ CP-even Higgs boson as shown in figure
\ref{annihilationchi}. The resulting constraint from the $\Omega
h^2_{\tilde{\chi}_1}$ observed limits as function of
$\tilde{\chi}_1$ mass,  for some selected regions in the parameter
space, using  $2 \sigma$ results reported by Planck satellite
\cite{Ade:2015xua}, is presented in figure \ref{MchiOGh2ALL1}
together with the LHC constraints, in particular, the SM-like
Higgs and gluino mass constraints.  Clearly from fig.
\ref{MchiOGh2ALL1}, the narrow range of the relic abundance limits
severely constrain this kind of DM candidates where only few
benchmark points are allowed.  However, the allowed points are
much larger than the corresponding ones in the MSSM. Recall that,
in the MSSM, no point with bino-like is allowed and much less
points for higgsino-like at very large $\tan \beta$ are allowed
\cite{Abdallah:2015hza}. Another remark is that the masses of
allowed $\tilde{\eta}_2$ lies in the range $100-1000$~GeV.

\section{Right-handed sneutrino: Scalar Dark Matter}
\vskip 0.2cm

We turn now to a scenario in which the lightest right-handed
sneutrino can serve as DM candidate.   To discuss this scenario,
we need first to  show the sneutrino mass matrix and discuss the
possibility of having lightest sneutrinos after diagonalization of
the mass matrix. To do this, we can write $\tilde{\nu}_{L}$,
$\tilde{\nu}_{R}$ and $\tilde{S}_2$ as \cite{Abdallah:2017gde} \be
\tilde{\nu}_L  = \frac{1}{\sqrt{2}} \left(\tilde{\nu}_L^+ + i\;
\tilde{\nu}_L^- \right),~~~~\tilde{\nu}_R  = \frac{1}{\sqrt{2}}
\left( \tilde{\nu}_R^+ + i\; \tilde{\nu}_R^-
\right),~~~~\tilde{S}_2 = \frac{1}{\sqrt{2}} \left( \tilde{S}_2^+
+ i\; \tilde{S}_2^- \right), \ee consequently, the sneutrino mass
matrix can be written as

\be M_{\tilde{\nu}}^2 = \left(%
\begin{array}{cc}
{\cal M}^2_+ & 0 \\
0 &{ \cal M}^2_- \\
\end{array}%
\right)\ee

where, for $\tilde{g}=0$, the CP-even/odd (right/left) sneutrino
mass matrix is given by

\begin{eqnarray*} {\cal M}^2_{\pm}={\small\fontsize{9}{9}\selectfont{ \left(%
\begin{array}{ccc}
m_{\tilde{L}}^2+m_D^2+\frac{1}{2}(M_Z^2\cos 2\beta+M_{Z'}^2\cos 2\beta') & \pm m_D(A_{\nu}+\mu\cot\beta) & m_D M_R\\
\pm m_D(A_{\nu}+\mu\cot\beta) & m_{\tilde{\nu}_R}^2+m_D^2+M_R^2-\frac{1}{2}M_{Z'}^2\cos 2\beta' & \pm M_R (A_S+\mu'\cot\beta')\\
m_D M_R & \pm M_R (A_S+\mu'\cot\beta') &
m_{\tilde{S}}^2+M_R^2+M_{Z'}^2\cos 2\beta'\\ \end{array}%
\right) }}\end{eqnarray*}

\begin{figure}[t]
\begin{center}
\vskip -0.5cm
\includegraphics[scale=0.7]{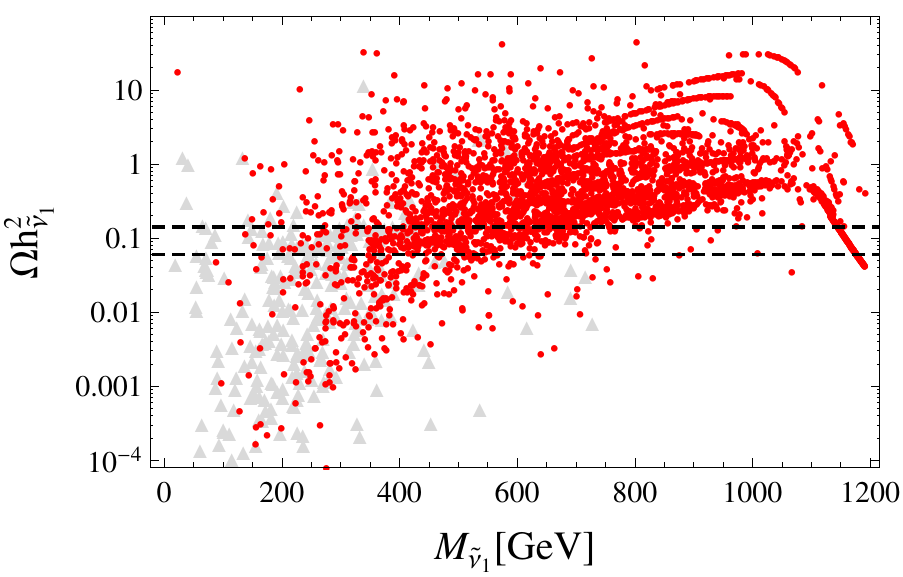}~~
\includegraphics[scale=0.7]{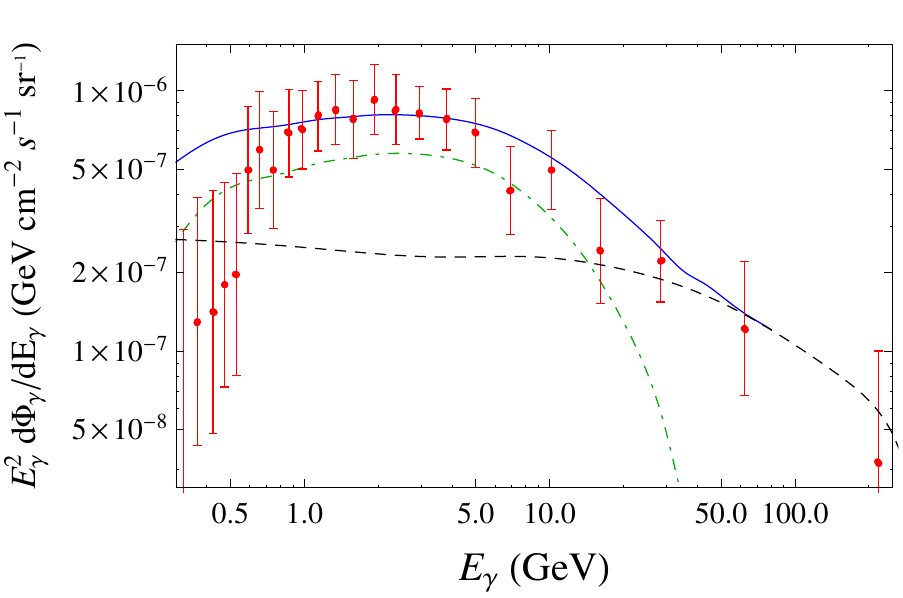}
\caption{Left, The thermal relic abundance of right-handed
sneutrino LSP as a function of its mass. The gray triangles denote
to the excluded points due to LUX upper bound. Horizontal lines
correspond to the Planck limits on DM abundance. Right, The
measured spectrum of gamma-rays within the ROI $2^{\circ} \leq |b|
\leq 20^{\circ}$ and $|l| \leq 20^{\circ}$ of the GC. The dashed
line shows the backgrounds. (Left panel) The gamma-rays spectrum
produced for the lightest sneutrino DM annihilation into
$WW~(91~\%)$ with $M_{\tilde{\nu}_1} \simeq 80.3$~GeV and total
annihilation cross section $\langle \sigma^{\rm ann} v \rangle
\simeq 2\times 10^{-26}~ {\rm cm}^3~{\rm s}^{-1}$ (the dot-dashed
green curve). The solid blue curve shows the sum of the signal and
its backgrounds.} \label{FERMILAT80GWW680MO}
\end{center}
\end{figure}

The diagonalization of $ {\cal M}^2_{\pm}$ is not straight forward
and can be only be done numerically. It turns out that the mass of
the lightest CP-odd sneutrino, $\tilde{\nu}^-_{i}$, is almost
equal to the mass of the lightest CP-even (right) sneutrino,
$\tilde{\nu}^+_{i}$ and can be of order ${\cal O}(100)$~GeV when
$\mu'$ and/or $A_S$ are of order $m_{\tilde{\nu}_R}$ and $M_R$,
i.e., $\sim {\cal O}(1)$~TeV. The lightest sneutrino
$\tilde{\nu}_1$  can be written in terms of $\tilde{\nu}^+_L$,
$\tilde{\nu}^+_R$, $\tilde{S}^+_2$ (in case of it is CP-even) as
\be \tilde{\nu}_1 = \sum_{i=1}^3 R_{1i} (\tilde{\nu}^+_L)_i +
\sum_{j=1}^3 R_{1j} (\tilde{\nu}^+_R)_j
 +  \sum_{k=1}^3 R_{1k} (\tilde{S}^+_2)_k.
\label{Gamma} \ee

with $R_{1j}=R_{1k}=\frac{1}{\sqrt{2}}$ for $j=k=1$ and the rest
of the $R$ coefficients are zeros which indicates that the
lightest sneutrino is a combination of $\tilde{\nu}^+_R$ and
$\tilde{S}^+_2$ and hence is mainly right-handed. The dominant
annihilation channels of $\tilde{\nu}_1$ are those annihilations
to CP-even Higgs bosons, $W^+ W^-$, $Z Z$ and three
light-neutrinos, $\nu^i_L$. The allowed range of the right-handed
sneutrino DM after taking into account the the  observed limits on
DM abundance, the Higgs mass and gluino mass constraints is shown
in figure \ref{FERMILAT80GWW680MO}. We note from that figure that,
the allowed values of $M_{\tilde{\nu}_1}$ range from $80$~GeV to
$1.2$~TeV. In fact this result leave a possibility of having DM
candidates in SUSY  due to the stringent restraints imposed on the
MSSM and in the BLSSM with neutralino DM candidates.

DM signals such as gamma-rays possibly produced by DM annihilation
can be probed using gamma-ray telescopes. With the ability to
search in energy range from $20$~MeV to $300$~GeV, the Large Area
Telescope (LAT) on the Fermi Gamma-ray Space Telescope (FGST)
mission can be considered a good tool for such probes. Galactic
Center (GC) gamma-ray photons excess in the range 3--4~GeV was
observed by Fermi-LAT
\cite{Daylan:2014rsa,Ackermann:2015zua,Hooper:2013rwa}. DM
particle with mass $\lesssim {\cal O}(100)$~GeV and annihilation
cross section of order $\langle \sigma^{\rm ann} v \rangle \simeq
10^{-26}~{\rm cm}^3~{\rm s}^{-1}$ can be one of the possible
sources responsible for this observed gamma-ray excess. Regarding
the $B-L$ neutralinos as DM candidates, the relic abundance
constrained their masses to be larger than $100$~GeV. In addition,
their annihilation cross sections in the galactic halo are of
order $10^{-30}~ {\rm cm}^3~{\rm s}^{-1}$ \cite{Abdallah:2017gde}.
As a consequence, they cannot account for such gamma-ray excess.
Tuning now to the lightest right-handed sneutrino as a candidate
of DM, based on the discussion in Ref.\cite{Abdallah:2017gde}, it
was shown that, see figure \ref{FERMILAT80GWW680MO}, right-handed
sneutrino with mass O(100) GeV  annihilating to $W^+ W^-$ bosons
can account for the observed gamma-ray excess. This can be
explained as the right-handed sneutrino is a scalar DM with
$s$-wave contribution to the annihilation cross section leading to
a value in galactic halo, almost equal to its value at the
decoupling limit, $\sim 10^{-26}~{\rm cm}^3~{\rm s}^{-1}$.

The effective scalar interactions of the DM, either being $B-L$
neutralino $\tilde{\chi}_1$ or the lightest Right-handed Sneutrino
 $\tilde{\nu}^{\rm R}_1$, with the up and down quarks can be expressed as
\be%
{\cal L}_{\text{scalar}} = f_q \overline{\tilde{\chi}}_1 \tilde{\chi}_1 \, \bar{q} q, %
\label{scalar}
\ee%
It is mainly due to $Z'$ exchange in the case of $B-L$ neutralino
while in the case of  $\tilde{\nu}^{\rm R}_1$ it is due to CP-even
Higgs bosons ($h$ and $h'$) exchanges. The $\tilde{\chi}_1$
coupling to protons and neutrons  are proportional to $f_u$ and
$f_d$. They are quite suppressed since $f_q \propto 1/M_{Z'}^2$,
with $M_{Z'}> 2$~TeV and hence the spin-independent cross section
of this scattering is expected to be very small. Regarding
 $\tilde{\nu}^{\rm R}_1$ coupling to protons and neutrons, the effective
coupling $f_q$ in eq.~(\ref{scalar}) is given by \be f_q \simeq
\frac{g_{\tilde\nu_1 \tilde\nu_1 h}~g_{q\bar{q} h}}{m_{h}^2} +
\frac{g_{\tilde\nu_1 \tilde\nu_1 h'}~g_{q\bar{q} h'}}{m_{h'}^2},
\label{fq} \ee

\begin{figure}[t]
\begin{center}
\vskip -0.5cm
\includegraphics[width=7cm,height=5.5cm]{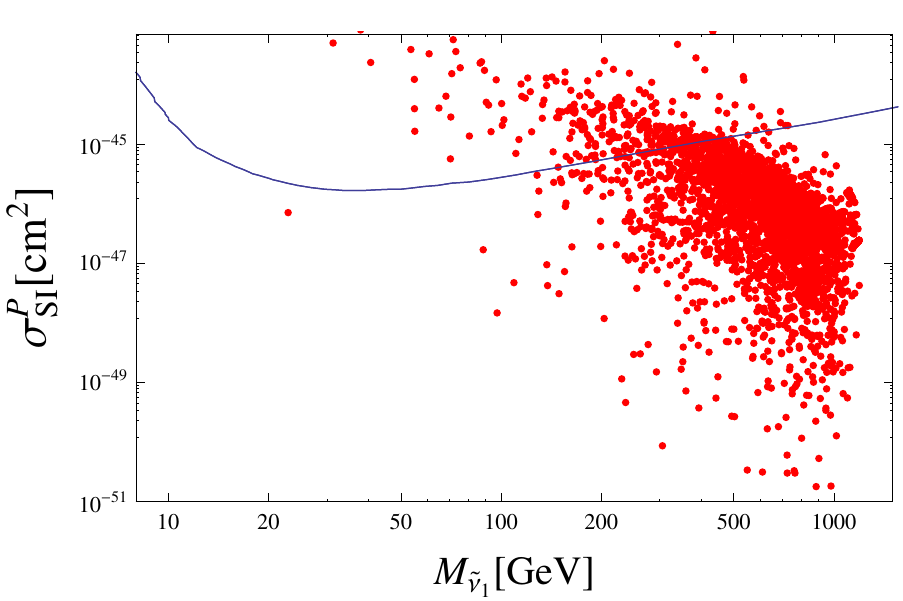}
\caption{The spin-independent cross section of the scattering
between the right-handed sneutrino LSP and proton versus its mass.
The blue curve is the recent LUX result.} \label{MsnuOMGh2snuLUX}
\end{center}
\vskip -0.5cm
\end{figure}

Based on the estimations in Ref.\cite{Abdallah:2017gde}, $f_q$ is
dominated by $h$ exchange as the effective coupling in this case
is of order ${\cal O}(10^{-3})~{\rm GeV}^{-1}$ compared to the
tiny one, ${\cal O}(10^{-7})~{\rm GeV}^{-1}$, in case of $h'$
exchange. It turns out that, the effective coupling of
$\tilde{\nu}^{\rm R}_1$ to proton and neutrino, is about three
order of magnitudes larger than the corresponding one in the case
of neutralinos. As a consequence, one would expect a larger
spin-independent cross section for sneutrino DM that may even
exceed the LUX limits as shown in Fig.\ref{MsnuOMGh2snuLUX}.

\section{Concluding remarks}

In this review, we presented a brief introduction related to the
formulation of the SM as the most successful theory, up to date,
in describing, explaining and predicting lot of well confirmed
experimental results. Due, to the failure of the SM in addressing
many issues, discussed in the text before, we highlighted the
success of SUSY, as one of the more popular candidates of physics
beyond SM, in providing solutions to the addressed problems in the
SM. These include, solution for the naturalness problems of the
Higgs sector in the SM, unification of the gauge couplings,  and
providing viable candidate fro cold dark matter. We showed also
that, SUSY provide a natural mechanism for understanding Higgs
physics and electroweak symmetry breaking and the inclusion of
gravity. With the running of LHC and with the help of future
experiments, different SUSY scenarios and parameter space can be
probed with a hope to a better understanding of the theory.

\end{document}